\documentclass[preprint2]{aastex}
\newcommand{\Lpar}{L_{*}}
\begin{document}
\title{The Fate of Sub-Micron Circumplanetary Dust Grains I: Aligned Dipolar Magnetic Fields}

\author{Daniel Jontof-Hutter\altaffilmark{1} and Douglas P. Hamilton\altaffilmark{1}}
\affil{Astronomy Department, University of Maryland, College Park, MD 20742-2421}

\begin{abstract}
 We study the stability of charged dust grains orbiting a planet and subject to gravity and the electromagnetic force. Our numerical models cover a broad range of launch distances from the planetary surface to beyond synchronous orbit, and the full range of charge-to-mass ratios from ions to rocks. Treating the spinning planetary magnetic field as an aligned dipole, we map regions of radial and vertical instability where dust grains are driven to escape or crash into the planet.  We derive the boundaries between stable and unstable trajectories analytically, and apply our models to Jupiter, Saturn and the Earth, whose magnetic fields are reasonably well represented by aligned dipoles. 
\end{abstract}
\section{Introduction}
The discoveries of the faint dusty ring systems of the giant planets beginning in the late 1970s greatly changed our understanding of planetary rings. Unlike Saturn's classical rings, which are most likely ancient \citep{can10}, dusty rings are young and are continually replenished from source satellites. Individual ring particles have short lifetimes against drag forces and other loss mechanisms, and because dusty rings are so diffuse, they are essentially collisionless. Furthermore, dusty rings are affected by a host of non-gravitational forces including solar radiation pressure and electromagnetism, which can sculpt them in interesting ways. 

Since the giant planets are far from the Sun and dusty rings are normally near their primary, radiation pressure is usually a weak perturbation to the planet's gravity. The electromagnetic force arising from the motion of charged dust grains relative to the planetary magnetic field, however, can be quite strong. In particular, with nominal electric charges, dust grains smaller than a fraction of a micron in radius are more strongly affected by electromagnetism than gravity. 

Dust in space acquires electric charges in several ways. Moving through the plasma environment produces a negative charge on a grain, since the plasma electrons are much lighter and swifter than ions and hence are captured more frequently by orbiting dust grains \citep{goe89}. On the other hand, sunlight ejects photo-electrons from the surface of a grain, and can cause positive charges \citep{hhm88}. Electron or ion impacts will also produce secondary electron emission, which also favors a net positive equilibrium charge on the grain \citep{whi81}. These currents interact in complicated ways; the charging of a dust grain depends on the physical properties of the grain itself and also on its charge history \citep{mv82}. \citet{gra08} provide an excellent review of these processes.

Many authors have investigated detailed aspects of the motion of charged grains in planetary magnetic fields, but no study has yet determined the orbital stability of grains for all charge-to-mass ratios launched at all distances in a systematic way. In this paper we explore the local and global stability of both positive and negative dust grains launched from ring particle parent bodies which themselves orbit at the local Kepler speed. 
\subsection{Motion in the Kepler and Lorentz Limits}
As grains with radii greater than several microns have small charge-to-mass ratios, electromagnetic effects are weak, and the grains orbit the planet along nearly Keplerian ellipses. In the frame rotating with the mean motion of the dust particle, the orbits appear as retrograde elliptical epicycles with a 2:1 aspect ratio \citep{mhh82}. When gravity acts alone, the vertical, radial and azimuthal motions all have precisely the same frequency. Equations governing the slow changes to the ellipse's orbital elements due to weak electromagnetic perturbations from a rotating aligned dipole magnetic field are given by \citet{ham93a}. These equations show that the three frequencies diverge slightly and are functions of the sign and magnitude of the charge as well as the distance from the planet and synchronous orbit.

Conversely, the very smallest dust grains approach the Lorentz limit, where the electromagnetic force dominates over gravity. In this regime, the frequencies of radial, vertical and azimuthal motions differ significantly. The radial oscillation is fastest and, as the electromagnetic force is perpendicular to the rotating magnetic field, particles gyrate about local field lines on typical timescales of seconds for dust, and microseconds for ions.

Dust grains typically oscillate vertically on a timescale of hours to days. Since this timescale is far slower than gyration, an adiabatic invariant exists and can easily be found. In the absence of forces other than electromagnetism, and ignoring planetary rotation, a dust grain's speed $v$ remains constant: $v^2 = v_{\perp}^2 + v_{\parallel}^2$, where $v_{\perp}$ and $v_{\parallel}$ are the speeds perpendicular and parallel to the magnetic field lines, respectively. The $v_{\perp}$ component determines the radius of the gyrocycle, while the $v_{\parallel}$ component moves the center of gyration to regions of differing magnetic field strength. If changes to a non-rotating magnetic field $\vec{B}$ are small over the size and time scales of gyromotion, the ratio $v^2_{\perp}/B$, where $B$ is the local strength of the field, is an adiabatic invariant \citep{dpl10} and hence is nearly constant. These two conditions provide an important constraint on the grain's motion parallel to the field lines. As a grain with a vertical velocity component climbs up a magnetic field line away from the equatorial plane, the field strength $B$ increases, $v_{\perp}$ also increases, and hence $v_{\parallel}$ must decrease. There is thus a restoring force towards the equatorial plane where the magnetic field strength is a local minimum, and the motion parallel to the field lines takes the form of bounce oscillations between mirror points north and south of the equator \citep{sto55}. \citet{tva80} studied the bounce motion of particles in the Lorentz limit at Saturn. Their results neglected the effects of planetary rotation, and hence are most applicable to slow rotators like Mercury and potentially some planetary satellites.

Finally, on the longest timescales (days), particles drift longitudinally with respect to the rotating magnetic field \citep{dpl10}, forced by a number of effects including gravity, the curvature of the magnetic field, and $\nabla B$. Because these motions are usually slow compared to the gyration and bounce frequencies, it is often useful to assume that in the Lorentz limit, grains are tied to the local field lines.

\subsection{Dust Affected by both Gravity and Electromagnetism}
For a broad range of grain sizes from nanometers to microns, both gravity and the Lorentz force are significant, and their combined effect causes a number of dynamical phenomena that are distinct from either limiting case. As dust in this size range predominates in many planetary rings (\citealt{bur99,dep99,sho08,kru09}), their dynamics have attracted much attention.

\citet{sb94} provide a general framework for the motion of dust started on initially Keplerian orbits. Since the radial forces on a dust grain at launch are not balanced as they are for a large parent body on a circular orbit, these dust grains necessarily have non-zero amplitude epicyclic motion. For the magnetic field configurations of the giant planets, a negatively-charged dust grain gyrates towards synchronous orbit while positively-charged dust initially moves away from this location. In fact, some positively-charged grains are radially unstable and either crash into the planet if launched inside synchronous orbit, or are expelled outwards if launched from beyond this distance. The latter have been detected as high-speed dust streams near Jupiter (\citealt{gru93,gru98}) and Saturn \citep{kem05}. Theoretical explanations for the electromagnetic acceleration process have been given by \citet{hmg93a,hmg93b}, \citet{hb93a} and \citet{gra00}.

\citet{mhh82} explored the shape and frequency of epicycles for negatively-charged grains in the transitional regime, where both EM effects and gravity are comparable. The epicycles make a smooth transition from perfectly circular clockwise (retrograde) gyromotion in the Lorentz limit, to 2:1 retrograde elliptical epicycles in the Kepler limit. \citet{mhh03} studied the shapes of epicyclic motion for positive grains and found that there is not a similarly smooth transition from prograde gyromotion to retrograde Kepler epicycles, and that the epicyclic motions of intermediately-sized grains cannot be represented as ellipses. The effects of gravity and electromagnetism compete for intermediate charge-to-mass values and motion can be primarily radial, leading to escape or collision \citep{hmg93a,hb93a}.

\citet{nh82,nh83a} and Northrop and Connerney (1987) studied the vertical motion of negatively-charged dust grains on circular uninclined orbits in a centered and aligned dipole field, a configuration most closely realized by Saturn. They found that some small grains on initially centrifugally-balanced circular trajectories inside the synchronous orbital distance are locally unstable to vertical perturbations, climbing magnetic field lines to crash into the planet at high latitudes. Some motions at high latitude, however, are stable: \citet{hhs99,hdh00} identified non-equatorial equilibrium points for charged dust grains, and showed than dust grains can orbit them stably. They characterized these ``halo'' orbits for positive and negative charged grains on both prograde and retrograde trajectories. \citet{hh01} used these analytical results to argue for a stable population of positively-charged grains in retrograde orbits and developed numerical models of such halo dust populations at Saturn. Grains that may populate these halos, however, are unlikely to result from the equatorial launches considered here.

If one of the dust grain's natural frequencies matches a characteristic spatial frequency of the rotating multipolar magnetic field, the particle experiences a Lorentz resonance \citep{bur85,sb87,sb92,hb93b,ham94}.  Lorentz resonances behave similarly to their gravitational counterparts and can have a dramatic effect on a dust grain's orbit, exciting large radial and/or vertical motions. These resonances have been primarily studied in the Kepler limit appropriate for the micron-sized particles seen in the dusty rings of Jupiter. In our idealized problem, with an axisymmetric magnetic dipole, Lorentz resonances cannot occur.

Variations in a dust grain's charge can also alter its trajectory over surprisingly rapid timescales. Gradients in the plasma properties, including density, temperature and even composition affect the equilibrium potential of a grain by altering the direct electron and ion currents. This can result in resonant charge variation with gyrophase, causing radial drift. Working in the Lorentz limit, \citet{nh83a} noted that with large radial excursions, the grain's speed through the plasma can vary significantly with gyrophase, leading to enhanced charging at one extremity. A similar effect occurs in the Kepler limit where resonant charge variation can cause a dramatic evolution in the orbital elements of a dust grain \citep{bs89}. \citet{nms89} found that the varying charge has a time lag that depends on the plasma density and grain capacitance. These time lags can cause grains to drift towards or away from synchronous orbit depending on the grain speed, and on any radial temperature or density gradients in the plasma. \citet{sb95} explored the effects of stochastic charging on extremely small grains, where the discrete nature of charge cannot be ignored. They found that Lorentz resonances are robust enough to survive  even for small dust grains with only a few electric charges.

The dynamics of time-variable charging may play an important role in determining the structure of Saturn's E ring \citep{jh04} and Jupiter's main ring and halo \citep{hj10}. Another example of charge variation occurs when the insolation of a dust grain is interrupted during transit through the planetary shadow. This induces a variation in charge that resonates with the grain's orbital frequency \citep{hb91}. \citet{hk08} found that this shadow resonance excites radial motions while normally leaving vertical structure unaltered. This effect can explain the appearance of the faint outward extension of Jupiter's Thebe ring, and the properties of its dust population sampled by the Galileo dust detector \citep{kru09}. 
\subsection{Research Goals}
In this study, we consider the orbits of charged grains launched in planetary ring systems. Our aim is to explore the boundaries between stable and unstable orbits in aligned and centered dipolar magnetic fields. Dipolar fields have the advantage of being analytically tractable while still capturing most of the important physics. Under what conditions are grains unstable to vertical perturbations? Which grains escape the planet as high speed dust streams? And which grains will strike the planet after launch? All of these instabilities depend on the launch distance of the grain and its charge-to-mass ratio. We first explore grain trajectories numerically and then derive analytical solutions for the stability boundaries that we find.

There are several standard choices for expressing the ratio of the Lorentz and gravitational forces. The charge-to-mass ratio $q/m$ in C/kg \citep{nh82} or in statCoulomb/g \citep{mhh03} may be the most straightforward, but it is cumbersome. For this reason, converting to the grain potential measured in Volts, which is constant for different-sized dust grains, is a common choice \citep{mhh82,sb94,hdh00,mhh03}. Yet another option is to express the charge-to-mass ratio in terms of frequencies associated with the primary motions of the grain, such as the gyrofrequency, orbital frequency and the spin frequency of the planet ($eg.$ \citealt{mhh82,mhh03}). 

We choose a related path, namely to fold $q/m$ and key planetary parameters into a single dimensionless parameter $\Lpar$ following \citet{ham93a}. Consider the Lorentz force in a rotating magnetic field:
\begin{equation}
\vec{F_{B}} = \frac{q}{c}(\vec{v}-\vec{\Omega}\times\vec{r})\times\vec{B},
\label{Bforce}
\end{equation}
where $c$ is the speed of light, $\vec{r}$ and $\vec{v}$ are the grain's position and velocity in the inertial frame, $\vec{\Omega}$ is the spin vector of the planet, and $\vec{B}$ is the magnetic field. We use CGS units here and throughout to simplify the appearance of the electromagnetic equations. The second component of Eq.~\ref{Bforce} is $q\vec{E}$, where $\vec{E}= -\frac{1}{c}(\vec{\Omega}\times\vec{r})\times\vec{B}$ is the so-called co-rotational electric field which acts to accelerate charged grains across magnetic field lines. Since a dipolar magnetic field obeys $\vec{B} = -g_{10}R_p^3/r^3 \hat{z}$ in the midplane (with $g_{10}$ the magnetic field strength at the planet's equator), $\vec{E}$, like gravity, is proportional to $1/r^2$ there. Thus the ratio of the electric force to gravity is both independent of distance and dimensionless:
\begin{equation}
\label{Lpar}
\Lpar = \frac{qg_{10} R_p^3 \Omega}{GM_pmc}.
\end{equation}
Here, $R_p$ and $M_p$ are the planetary radius and mass, $m$ is the dust grain mass and $G$ is the gravitational constant. Note that the the sign of $\Lpar$ depends on the product of two signed quantities, $q$ and $g_{10}$. For all of the giant planets, the magnetic north pole is in the northern hermisphere, and $g_{10} > 0$. However, for the Earth at the current epoch, $g_{10} < 0$ and the magnetic and geographic poles are in opposite hemispheres.

 We have made a slight notational change $L \rightarrow \Lpar$ from \citet{ham93a,ham93b} to avoid confusion with the L-shell of magnetospheric physics. Choosing $\Lpar$ as an independent variable takes the place of assuming a particular electric potential, grain size and grain density. We focus our study primarily on Jupiter, the planet with by far the strongest magnetic field, but also apply our results to Saturn and to the Earth.
\section{Numerical Simulations}
Approximating Jupiter's magnetic field as an aligned dipole by including just $g_{10} = 4.218$ Gauss \citep{des83}, we tested the stability of dust grain orbits over a range of grain sizes and launch distances both inside and outside synchronous orbit. We used a Runge-Kutta fourth-order integrator and launched grains at the local Kepler speed with a small initial latitude of $\lambda = 0.01^{\circ}$. This tiny nominal value ensures a launch close to the midplane, whilst avoiding potential numerical problems of launching a grain precisely at $\lambda = 0$. Non-zero launch speeds from the parent particle do have a small effect on the stability boundaries, one that we will explore in more depth in a future study.

 Our models treat the grain charge as constant and neglect $J_2$, other higher-order components of the gravitational field, and radiation pressure. For both negative and positive grains, we ran simulations for a grid of 80 values of $\Lpar$ and 100 launch distances ($r_L$). The charge-to-mass ratio spans four decades from the Lorentz regime where EM dominates ($|\Lpar| >>1$), to the Kepler regime where gravity reigns ($|\Lpar| << 1$). The range of launch distances extends from the planetary surface to well beyond the synchronous orbital distance ($R_{syn}$), and trajectories were followed for up to 0.1 years. With some experimentation, we determined that all relevant dynamical timescales are $< 0.1$ years and that for longer integration times, the appearance of our stability plots does not change significantly.

\begin{figure}[placement h]
\includegraphics [height = 2.1 in]{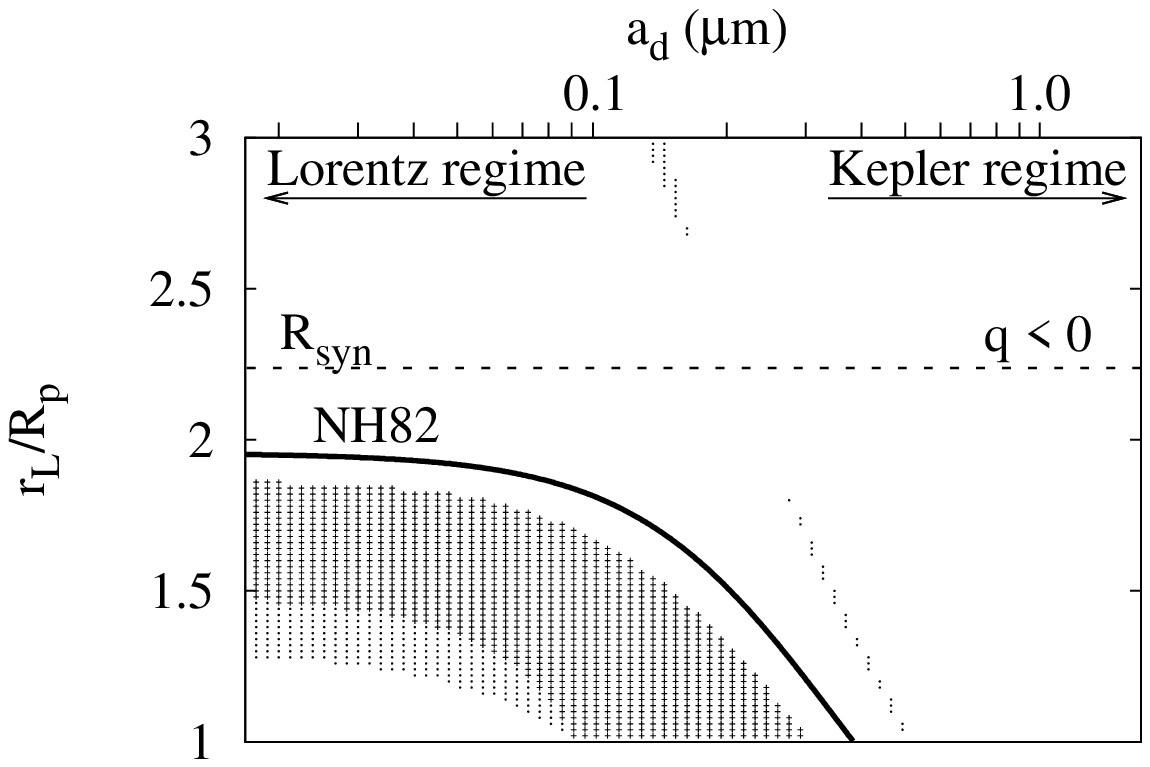}
\newline
\includegraphics [height = 2.1 in]{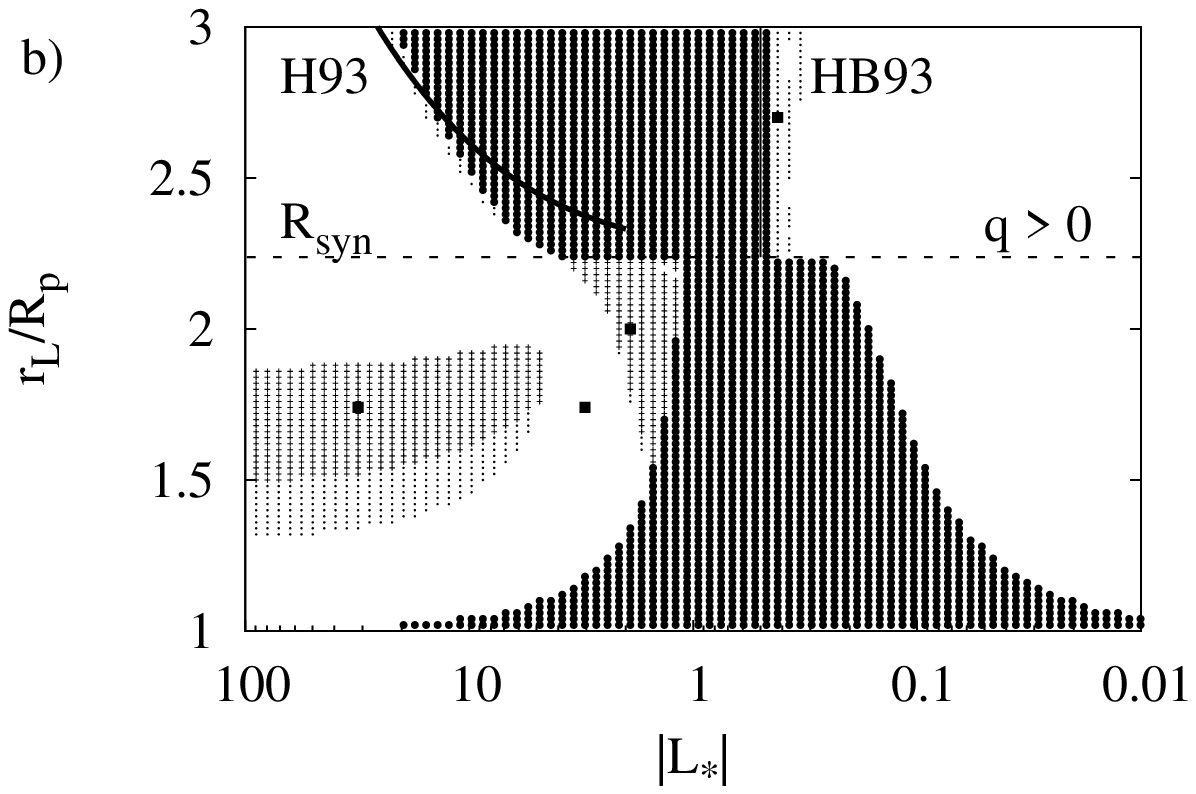}
\caption{Stability of Kepler-launched \textbf{a)} (negative) and \textbf{b)} (positive) dust grains at Jupiter. We model the planet with a spherically-symmetric gravitational field, and a centered and aligned dipolar magnetic field. All grains were launched with an initial latitude of $\lambda = 0.01^{\circ}$ and followed for 0.1 years.  The horizontal dashed line in both panels denotes the synchronous orbital distance at $R_{syn} = 2.24 R_p$. The grain radii ($a_d$) in microns along the upper axis are calculated assuming a density of 1 g/cm$^3$ and an electric potential of $\pm 5 V$ so that $|\Lpar|  = 0.0284/a_d^2$. Dust grains in the white regions and lightest grey areas survive the full 0.1 years, with the latter reaching latitudes $\lambda$ in excess of $5^{\circ}$. Grains in the moderately-grey areas are vertically unstable and strike the planet, also at high latitudes ($\lambda >5^{\circ}$). The darkest regions, seen only in panel \textbf{b}, are radially unstable grains that crash into the planet (those with $r_L<R_{syn}$), or escape to beyond $r_{esc} = 30R_p$ (from $r_L>R_{syn}$) at latitudes less than $5^{\circ}$. We overplot three analytically-derived stability boundaries, obtained by \citet{nh82} for negative grains, by \citet{hmg93a} for small positive grains, and by \citet{hb93a} for large positive grains. Each point on the plot is a trajectory, some of which (marked by filled squares), are illustrated in detail in Figs.~\ref{fig:traj1} to~\ref{fig:JUPg10randomwalk}.}
\label{fig:g10num} 
\end{figure}

In Fig.~\ref{fig:g10num} we plot the fate of 8000 negative and 8000 positive dust grains and find complex regions of instability. The negatively-charged dust grains in Fig.~\ref{fig:g10num}a display only vertical instability at moderate to high $\Lpar$ and inside $R_{syn}$. Some are bound by high latitude restoring forces (locally unstable, light grey) whilst others crash into the planet at high latitude (both locally and globally unstable, darker grey). To separate these globally stable grains from locally stable ones, we choose a latitude threshold at $\lambda_m = 5^{\circ}$. Although 5$^{\circ}$ is a small latitude, it is far greater than the launch latitude of 0.01$^{\circ}$; any grains excited beyond $\lambda_m$ are clearly locally unstable, and we determined that our results were fairly insensitive to actual value of $\lambda_m$.

 \citet{nh82} derived a boundary for the threshold between locally stable and unstable trajectories for negatively-charged dust and found that grains launched within a certain distance should leave the equatorial plane (NH82 curve in Fig.~\ref{fig:g10num}a). In the Lorentz limit, the vertical instability allows grains to climb up local magnetic field lines into regions of stronger magnetic field, while for smaller $\Lpar$ the path taken by these grains follows the lines of a pseudo-magnetic field which includes the effects of planetary rotation \citep{nh82}. The Northrop curve however, is not a good match to our data which reveal additional stable orbits (white areas) immediately inside this boundary and also close to the planetary surface. These differences arise from the fact that \citet{nh82} assumed that grains are launched at their equilibrium circular speeds, which differ from the circular speeds of parent bodies when $\Lpar \ne 0$. Conversely, we launch our grains at $v = \sqrt{G M_p/r}$, the circular speed of the parent body, which is appropriate for debris produced by cratering impacts into these objects. In section 5, we develop a vertical stability criterion appropriate for our launch conditions.

The situation for positive grains is quite different. Figure~\ref{fig:g10num}b shows a less extensive region of vertical instability than Fig.~\ref{fig:g10num}a, and one that is not active close to Jupiter. More dramatic, however, are two regions of radial instability (darkest grey areas), separated by the synchronous orbital distance. Grains inside $R_{syn}$ are driven to strike Jupiter, while those outside escape the planet. If grains move beyond $r_{esc} = 30R_p$, the inner magnetosphere, we consider them to have escaped. As with $\lambda_m$, our numerical results are fairly insensitive to the exact value chosen for $r_{esc}$, so long as it is large.

\begin{figure}[placement h]
\includegraphics [height = 2.1 in]{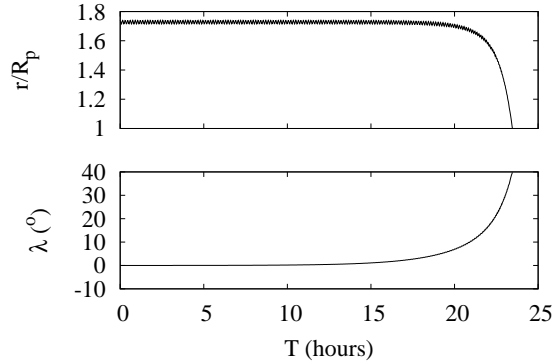}
\caption{The trajectory of a positively-charged grain orbiting Jupiter after launch at $r_L = 1.74 R_p$, with $\Lpar=31.31$ ($a_d = 0.03\mu$m). We plot the scaled distance and latitude of the dust grain against time. The small, rapid radial gyration is just visible in the upper plot. The dust grain is vertically unstable on a much longer timescale and ultimately crashes into the planet. This trajectory is the left-most filled square in Fig.~\ref{fig:g10num}b.}
\label{fig:traj1} 
\end{figure}

\begin{figure}[placement h]
\includegraphics [height = 2.1 in]{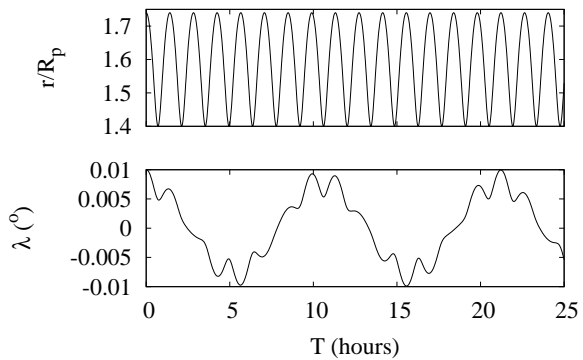}
\caption{The trajectory of a stable positively-charged grain orbiting Jupiter after launch at $r_L = 1.74 R_p$, with $\Lpar=3.04$ ($a_d = 0.097 \mu$m). The grain undergoes radial oscillations much larger than in Fig.~\ref{fig:traj1} but its latitude remains low. Here the bounce period is $\sim 7$ times longer than the gyroperiod.}
\label{fig:stablechannel} 
\end{figure}

To characterize the individual trajectories that make up Fig.~\ref{fig:g10num}, we explore a few examples in detail, focusing on the positively-charged dust grains and proceeding from smaller to larger grains. Figure~\ref{fig:traj1} shows the trajectory of a dust grain that becomes vertically unstable and crashes into the planet at high latitude. These smallest grains spiral up magnetic field lines, which for a dipole are given by $r/\cos^2\lambda = r_L$ \citep{dpl10}; collision with the planet or reflection from a high latitude mirror point typically occurs within a few tens of hours. By contrast, Fig.~\ref{fig:stablechannel} shows an electromagnetically-dominated grain that remains stable at low latitude.

\begin{figure} [placement h]
\includegraphics [height = 2.1 in] {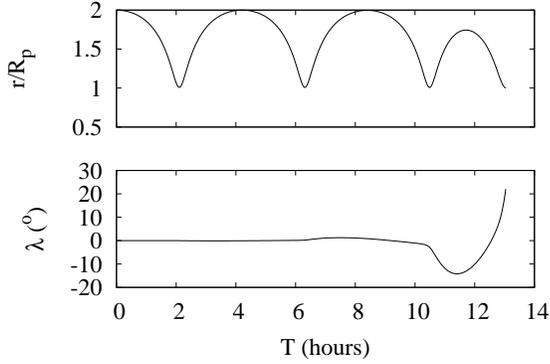}
\caption{The trajectory of a positive grain inside $R_{syn}$ ($r_L = 2.0R_p, \Lpar= 1.908$, $a_d = 0.122 \mu$m). Here, unlike Fig.~\ref{fig:stablechannel}, large radial motions ultimately excite vertical motions, forcing the trajectory to end with a collision at the planetary surface after just a few orbits.}
\label{fig:JUPg10resunstable}
\end{figure}

A more subtle interplay between radial and vertical motions is illustrated in Fig.~\ref{fig:JUPg10resunstable}. This grain is outside the radial instability region in which grains collide with the planet at low latitude (darkest grey).  Instead, large radial motions lead to instability in the vertical direction, and ultimately, the grain strikes the planet at high latitude. Notice the two white dots near ($\Lpar = 1.34$, $r_L/R_p = 2.2$) in Fig.~\ref{fig:g10num}b, signifying grains that survive the full 0.1 year integration. These trajectories are indeed stable (for at least 100 years) and, as the effect is much more prominent for the Earth, we discuss it in more detail in section 6.

\begin{figure} [placement h]
\includegraphics [height = 2.1 in] {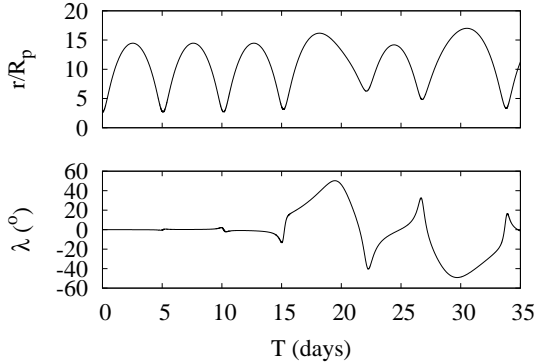}
\caption{The trajectory of a positive grain outside $R_{syn}$ ($r_L = 2.7R_p, \Lpar = 0.419$, $a_d = 0.26 \mu$m). As in Fig.~\ref{fig:JUPg10resunstable}, large radial oscillations eventually excite large vertical oscillations. Since the dust grain has $\Lpar <\frac12$, it is energetically required to remain bound \citep{hb93a}. Here $T$ is measured in Earth days.}
\label{fig:JUPg10randomwalk}
\end{figure}

Finally, Fig.~\ref{fig:JUPg10randomwalk} shows a dust grain just inside the \citet{hb93a} $\Lpar = \frac12$ stability limit. Although the dust grain does not escape, the non-linearity of its radial oscillation is large enough to excite substantial vertical motions. 

A glance at Fig.~\ref{fig:g10num} shows that most stability boundaries are unexplained. The \citet{nh82} vertical stability boundary does not match the numerical data especially well, and only applies to negative grains. For positive grains, \citet{hmg93b} provided an approximate criterion for radial escape, which they applied far from synchronous orbit near Io. Their criterion is based on a comparison between the radius of gyromotion $r_g$, and the length scale over which the magnetic field changes substantially, namely where $|B/(r_g\nabla B)| \approx 10$, with the gyroradius calculated in the Lorentz limit. Although not intended for use near synchronous orbit where $r_g \rightarrow 0$, we nevertheless plot it on the left side of Fig.~\ref{fig:g10num}b. Finally, the \citet{hb93a} $\Lpar = \frac12$ limit, derived from an energy argument, is a good match to the largest escaping grains. There is however, no analytical model for the broad class of grains that strike the planet. Accordingly, we seek to develop a unified theory that can cleanly determine all of these boundaries. We take up this task first for radial and then for vertical motions.
\section{Local Radial  Stability Analysis}
Consider a centered magnetic dipole field that rotates with frequency $\Omega$ around a vertical axis aligned in the $z$-direction. \citet{nh82} derived the Hamiltonian for a charged dust grain in the rotating frame in cylindrical coordinates:
\begin{equation}
\label{nh82ham}
H = U(\rho,z) + \frac{\dot{\rho}^2+\dot{z}^2}{2}
\end{equation}
where $\dot{\rho}$ and $\dot{z}$ are the radial and vertical velocity components. The potential is given by
\begin{eqnarray}
\label{potentialcylinder} 
U(\rho,z) = \frac{1}{2\rho^2}\left(\frac{p_{\phi}}{m}-\frac{GM_p\rho^2\Lpar}{\Omega r^3}\right)^2+ &\nonumber  \\ \frac{G M_p}{r}\left(\frac{\Lpar \rho^2}{r^2}-1\right)
\end{eqnarray}
where the spherical radius $r$ satisfies $r^2 = \rho^2 +z^2$ \citep{nh82,sb94,hdh00,mhh03}. Equation~\ref{potentialcylinder} is the sum of two energetic components: first the azimuthal specific kinetic energy, which can be expressed as a function of $r$ using the conservation of angular momentum, and then the potential associated with both the corotational electric field and gravity.  Note that we have chosen the zero of our potential to be approached as $\rho\rightarrow \infty$. Because $U(\rho,z)$ is independent of $\phi$, the azimuthal coordinate, the canonical conjugate momentum $p_{\phi}$ is a constant of the motion. For our launch condition from a large parent body on a circular orbit at $r = r_L$:
\begin{equation}
\label{Pphi}
\frac{p_{\phi}}{m} = r_L^2(n_L+\Omega_{gL})
\end{equation}
\citep{sb94}, where $n_L$ and $\Omega_{gL}$ are the Kepler frequency and gyrofrequency evaluated at the launch distance $r_L$:
\begin{equation}
n_L = \sqrt{\frac{GM}{r_L^3}},
\label{nkepL}
\end{equation}
and
\begin{equation}
\label{gyrofrequencyL}
\Omega_{gL} = \frac{qB}{mc} = \frac{n_L^2\Lpar}{\Omega}. 
\end{equation}
Notice that in the gravity limit ($\Lpar \rightarrow 0$), Eq.~\ref{Pphi} reduces to $r_L^2n_L$, the specific angular momentum about the planet, while in the Lorentz limit ($\Lpar \rightarrow \infty$), it is $r_L^2\Omega_{gL}$, the specific angular momentum about the center of gyromotion that moves with the magnetic field. 

If the motion of the particle is radially stable, it exhibits epicyclic motion about an equilibrium point determined from Eq.~\ref{potentialcylinder}. The existence of equilibrium points requires that $\frac{\partial U}{\partial \rho} = \frac{\partial U}{\partial z} = 0$, both in the equatorial plane \citep{nh82} and at high latitudes (\citealt{hhs99,hdh00}). The local stability of the equilibrium points, defined as whether oscillations about these points remain small, is then determined by considering the second derivatives of the potential. Given our launch condition, we focus on the equatorial equilibrium points which are of greatest interest. For these, $\frac{\partial^2 U}{\partial \rho \partial z} \bigg{|}_{\rho= \rho_c, z=0} = 0$, $r \rightarrow \rho$, and radial and vertical motions are initially decoupled and may be considered separately (\citealt{nh82,mhh03}). 

The equilibrium point is the guiding center of epicyclic motion. Grains launched at the guiding center have canonical conjugate momenta that are different from our Kepler-launched grains: namely, $\frac{p_{\phi}}{m} = \rho_c^2(\omega_c+\Omega_{gc})$, where $\omega_c$ is the orbital frequency of a grain at the guiding center, $\Omega_{gc}$ is the gyrofrequency at the guiding center, and $\rho_c$ is the guiding center distance in the equatorial plane. A local radial stability analysis is most relevant for our Kepler-launched grains if an equilibrium point is not too distant. Accordingly, it is important to distinguish between quantities evaluated at the Kepler launch position and those determined at the guiding center. Here and throughout, we use the subscript $c$ for the guiding center and the subscript $L$ for the launch position. At the equilibrium point, $\frac{\partial U}{\partial \rho}\bigg{|}_{\rho= \rho_c, z=0} = 0$, which evaluates to:
\begin{equation}
\omega_c^2\rho_c + \frac{GM_p\Lpar}{\rho_c^2} \left(1-\frac{\omega_c}{\Omega}\right)-\frac{GM_p}{\rho_c^2} = 0.
\label{dUdr}
\end{equation}
Physically, Eq.~\ref{dUdr} just implies a balance of forces in the rotating frame, whereby the centrifugal force, the Lorentz force and gravity sum to zero. We solve Eq.~\ref{dUdr} for the angular speed of the guiding center $\omega_c$, and find two real roots for $\Lpar <1$, which includes all negative charges. For $\Lpar >1$ conversely, two equilibrium points exist only if
\begin{equation}
\frac{\rho_c^3}{R_{syn}^3} \le \frac{\Lpar^2}{4(\Lpar-1)}.
\label{noequilibrium}
\end{equation}
Two equilibria always exist inside $R_{syn}$ and everywhere for $\Lpar >> 1$ and $\Lpar <<1$. There are no equilibrium points in a region starting at ($\Lpar = 2$, $\rho_c = R_{syn}$) in Fig.~\ref{fig:g10num}b, and opening upward to include an increasing range of $\Lpar$ values for increasing distance $\rho_c$. In this region, no equilibrium point exists, and grains are guaranteed to be locally unstable. Not surprisingly, this region is fully contained within the unstable portion of Fig.~\ref{fig:g10num}b (darkest grey region outside $R_{syn}$). The existence of an equilibrium point, therefore, is a necessary prerequisite for stability. 

Additional instability in Fig.~\ref{fig:g10num}b comes from two sources:
i) the intrinsic instability of the equilibrium point, if it exists, and
ii) large amplitude motions about a locally stable equilibrium point. Large oscillations are beyond the scope of a local stability analysis and so we focus on small amplitude radial motion near an equilibrium point, which takes the form
\begin{equation}
\ddot{\rho}+\frac{\partial ^2U}{\partial \rho^2}\rho = 0.
\label{hookeslaw}
\end{equation}
Small radial motions are stable when  $\frac{\partial ^2U}{\partial \rho^2}\bigg{|}_{\rho= \rho_c, z=0} = \kappa_c^2  > 0$, which, from Eq.~\ref{potentialcylinder} can be written as:
\begin{equation}
\kappa_c^2 =  \omega_c^2-4\omega_c\Omega_{gc}+\Omega_{gc}^2
\label{kappa}
\end{equation} 
(\citealt{mhh82}; Northrop and Hill 1982; Mitchell \textit{et al.} 2003). Note here that the gyrofrequency $\Omega_{gc}$ is evaluated at the guiding center, and is given by Eq.~\ref{gyrofrequencyL} with the subscript change: $L \rightarrow c$. The epicyclic frequency $\kappa_c$ reduces to the Kepler orbital frequency $n_c$ at the guiding center $r_c$ in the gravity limit, and to the gyrofrequency $\Omega_{gc}$ in the Lorentz limit. Radial excursions in both of these cases are small and, since $\kappa_c^2 > 0$, are guaranteed to be stable. 

Radial motions are also initially small near synchronous orbit where electromagnetic forces are very weak (Eq.~\ref{Bforce}), and so a local stability analysis is also applicable. At synchronous orbit, $\omega_c = n_c = \Omega$ and Eq.~\ref{kappa} reduces to $\kappa_c^2 = \Omega^2(1-4\Lpar+\Lpar^2)$, which is positive for small or large $\Lpar$. For $2-\sqrt{3} < \Lpar < 2+\sqrt{3}$, however, Eq.~\ref{hookeslaw} shows that radial motions near synchronous orbit are locally unstable. Comparing this analysis with Fig.~\ref{fig:g10num}b, we see that all orbits with $r_L \sim R_{syn}$ that are locally stable are, not surprisingly, also globally stable. The converse, however, does not hold: although most of the locally unstable orbits are also globally unstable, some are in fact globally stable (\textit{e.g}. $\Lpar < \frac12$ just outside $R_{syn}$ in Fig.~\ref{fig:g10num}b). In conclusion, the local analysis is consistent with our numerical experiments but cannot fully account for our stability boundaries. Accordingly, we turn to a global analysis, pausing first to put the potential of Eq.~\ref{potentialcylinder} into a more useful form and to derive the radius of gyration, $r_g$.
\subsection{Radius of Gyration}
\begin{figure}[placement h]
\includegraphics [height = 2.1 in]{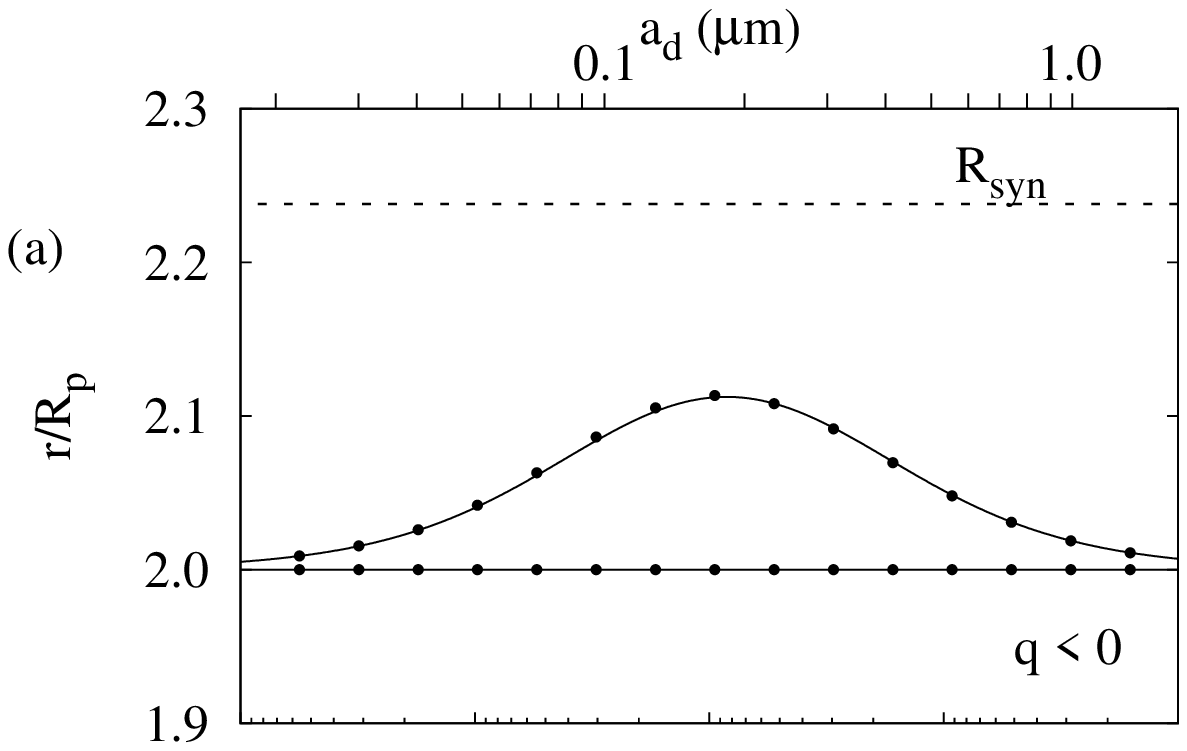}
\newline
\includegraphics [height = 2.1 in]{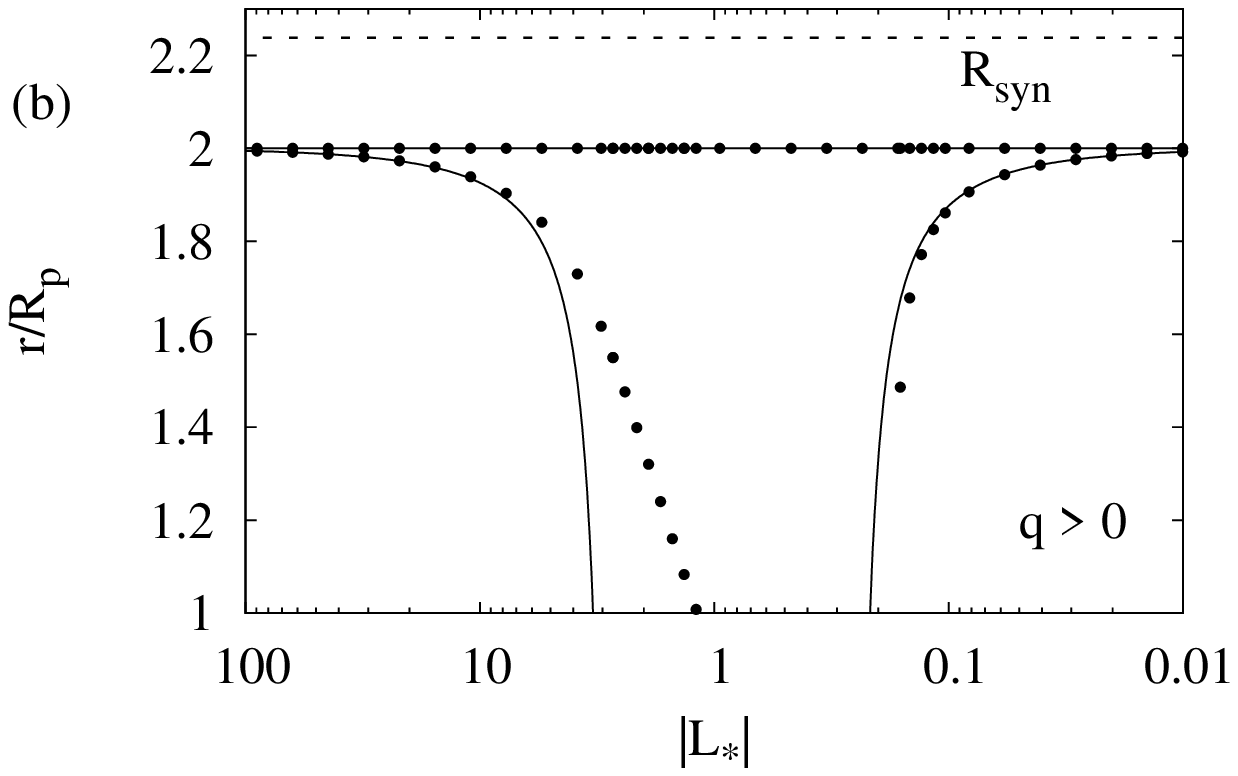}
\caption{The radial range of \textbf{(a)} negative and \textbf{(b)} positive grains launched azimuthally with the Kepler speed $v = \sqrt{GM/r_{L}}$ at $2.0R_p$. Both numerical data (points) and the analytical results (curves from Eq.~\ref{gyroradius}) are included. The total radial excursion is twice the epicyclic radius $r_g$.}
\label{fig:radrange} 
\end{figure}

With our launch condition, grains are often far enough from an equilibrium point that the small oscillation approximation of Eq.~\ref{hookeslaw} is invalid. This is particularly true far from $R_{syn}$ and for $\Lpar \approx 1$. Returning to the effective potential of Eq.~\ref{potentialcylinder} with the canonical conjugate momentum determined by launching the grain at the Kepler speed (Eq.~\ref{Pphi}), and limiting our attention to planar orbits for which $z = 0$ and $r = \rho$, we express the potential as a quartic polynomial function of distance and a quadratic function of $\Lpar$:
\begin{eqnarray}
\label{polynomialpotential1}
U(r,\Lpar) = \frac{GM_p}{r_L}\left(A\frac{r_L^4}{r^4}+B\frac{r_L^3}{r^3}+C\frac{r_L^2}{r^2}+D\frac{r_L}{r}\right),
\end{eqnarray}
with dimensionless coefficients
\begin{eqnarray*}
\label{polynomialpotential2}
A&=& \frac{n_L^2 \Lpar ^2}{2\Omega^2} \\
B&=&-\frac{n_L\Lpar}{\Omega}\left(\frac{n_L\Lpar}{\Omega}+1\right)\\ 
C&=&\frac12\left(\frac{n_L\Lpar}{\Omega}+1 \right)^2 \\
D&=&\Lpar-1. 
\end{eqnarray*}
To determine the radius of the epicycles ($r_g$) induced by a Kepler launch, we follow the procedure of \citet{sb94}, and solve for the distance to the potential minimum where $\frac{\partial U}{\partial r} \bigg{|}_{\rho= \rho_c, z=0} = 0$. Note that this is only valid to first order in small quantities, since we are effectively assuming that the potential is symmetric about the equilibrium point. Evaluating the derivative, multiplying by $r^5$, setting $r = r_L + r_g$, and assuming $r_g<< r_L$, we obtain the epicycle radius for a grain launched at $r_L$ in terms of parameters known at launch:
\begin{equation}
r_g = \frac{r_L(\Omega-n_L)\Omega_{gL}}{\Omega_{gL}^2-\Omega_{gL}(3\Omega+n_L)+n_L^2}.
\label{gyroradius}
\end{equation}  
In this limit, the radial range of motion of a dust grain is simply $2|r_g|$, and the grain reaches a turning point at $r_t = r_L+2r_g$. Note the sign conventions used here; $r_g$ and $\Omega_{gL}$ may be either positive or negative; thus negative grains (with $\Omega_{gL} < 0$) always gyrate towards $R_{syn}$. Equation~\ref{gyroradius} corrects a sign error in \citet{sb94} which led to an artificial disagreement between the numerical and analytical model in their Fig. 6. Equation~\ref{gyroradius}, by contrast, shows excellent agreement with our numerical data for negative grains (Fig.~\ref{fig:radrange}a). The peak in Fig.~\ref{fig:radrange}a, for oscillations towards synchronous orbit, occurs at
\begin{equation}
r_g = \frac{r_L}{3}\left(\frac{n_L-\Omega}{n_L+\Omega}\right), \Lpar = -\frac{\Omega}{n_L}.
\label{maxrange}
\end{equation}  
Equation~\ref{maxrange} predicts that grains with $\Lpar = -\frac{\Omega}{n_L}$ launched near $R_{syn}$ reach about halfway to the synchronous orbital distance, in agreement with Fig.~\ref{fig:radrange}a.

For the positive grains, Eq.~\ref{gyroradius} gives the proper the radial range about stable local minima in both the Lorentz limit and in the Kepler limit (Fig.~\ref{fig:radrange}b). At critical values of $\Lpar$, however, $|r_g| \rightarrow \infty$ and the assumptions under which Eq.~\ref{gyroradius} was derived are violated. This is readily apparent in the decreasing quality of the match between the theory and the data for intermediate-sized grains in Fig.~\ref{fig:radrange}b. Note that this is the same region where \citet{mhh03} find large non-elliptical gyrations.
 Nevertheless, the relatively close agreement between theory and numerical data in Fig.~\ref{fig:radrange} confirms that the epicyclic model is usually a good assumption in planetary magnetospheres.
\section{Global Radial Stability Analysis}
 Our local radial stability analysis makes a number of successful predictions, but cannot fully account for the boundaries in Fig.~\ref{fig:g10num}b, primarily because of the large radial excursions experienced by the positive grains. The quartic potential within the equatorial plane given by Eq.~\ref{polynomialpotential1} contains all the information necessary to determine which grains strike the planet and which escape into interplanetary space.
\subsection{Escaping Grains}
Close to the planet, the $A/r^4$ term of Eq.~\ref{polynomialpotential1} dominates, and $U(r\rightarrow 0,\Lpar) \rightarrow +\infty$, while for the distant particles we have $U(r\rightarrow\infty,\Lpar) \rightarrow 0$. Accordingly, the quartic potential can have at most three stationary points (one local maximum and two local minima). Setting $r=r_L$ gives a simple form for the launch potential
\begin{equation}
\label{Lhalfescape}
U(r_L,\Lpar) = \frac{GM_p}{r_L}\left(\Lpar-\frac12\right).
\end{equation}

Energetically, a particle is able to escape if $U(r_L,\Lpar) > U(r\rightarrow \infty, \Lpar) = 0$ and we immediately recover the $\Lpar < 1/2$ stability criterion of \citet{hb93a}. Note that only positive grains can escape from a dipolar magnetic field and that, in principle, grains with $\Lpar > \frac12$ at all launch distances, both inside and outside $R_{syn}$ are energetically able to escape. Whether or not they do so depends on the form of $U(r,\Lpar)$, in particular, on the possible existence of an exterior potential maximum with $U(r_{peak},\Lpar)>U(r_L,\Lpar)$. 

Analysis of Eq.~\ref{polynomialpotential1} shows that the potential prevents all grains launched with Kepler initial conditions from crossing $R_{syn}$. Positive grains gyrate away from $R_{syn}$, while negative grains cannot reach $R_{syn}$ (Eq.~\ref{gyroradius}, Fig.~\ref{fig:radrange}).

\begin{figure} [placement h]
\includegraphics [height = 2.0 in] {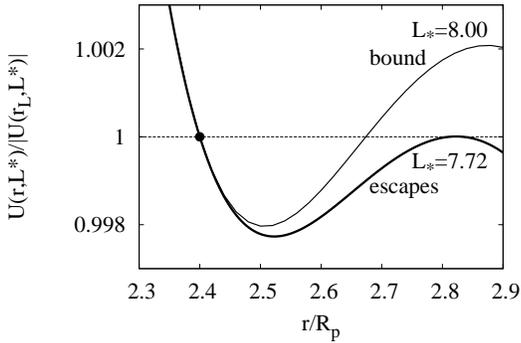}
\caption{Potential wells for positive grains launched just outside Jupiter's synchronous orbit, at $r_L = 2.4R_p$. For $\Lpar = 8.00$ ($a_d = 0.0596 \mu$m), a distant local maximum bounds the motions. If $\Lpar=7.72$ ($a_d = 0.0607 \mu$m) the distant peak in the potential is at the radial turning point, and the potential is equal to the launch potential; this is the stability threshold. For smaller $\Lpar$, the peak is lower and escape occurs.}
\label{fig:wells1}
\end{figure}

Outside $R_{syn}$, $U(r,\Lpar)$ monotonically decreases for $\Lpar \gtrsim \frac12$. Thus $\Lpar = \frac12$ is a global stability boundary and it matches Fig.~\ref{fig:g10num}b very well. For larger $\Lpar$ (smaller grains), the topography is illustrated in Fig.~\ref{fig:wells1}.
 Stability is determined by the height of the distant peak in the potential. For $\Lpar \sim 1$ no such peak exists. For larger $\Lpar$, however, the radial potential decreases with distance from $r_L$, then increases to the distant peak, and finally declines to zero as $r \rightarrow \infty$. 

Consider the quartic equation $U(r,\Lpar) - U(r_L,\Lpar) = 0$, which by construction, has one root at $r = r_L$, and one root at a more distant turning point $r = r_t$. The critical quartic, where the turning point is also a local maximum (as in Fig.~\ref{fig:wells1}) has a double root at $r = r_t$. By factoring out ($r-r_L$), and then differentiating with respect to $r$, we find a quadratic equation for the location of the turning point; $r_t$ varies smoothly from $r_t = r_L$ at synchronous orbit to $r_t = \frac32 r_L$ for $r_L >> R_{syn}$. The stability boundary, $r_L(\Lpar)$ starts at ($r=R_{syn}, \Lpar = 2+\sqrt{3}$) and asymptotes to
\begin{equation}
\label{Lthird}
\frac{r_L}{R_{syn}} = \left(\frac{2 \Lpar}{27}\right)^{\frac13}
\end{equation}
for $r_L >> R_{syn}$. Equation~\ref{Lthird} for $r>> R_{syn}$ is a useful approximation for the boundary far from $R_{syn}$, which nicely compliments the exact value we have found at the synchronous orbital distance. The full solution for the boundary $r_L(\Lpar)$ is given by a rather messy cubic equation and so we resort to numerical methods for its solution, which we plot on Fig.~\ref{fig:g10JUP}b.  
\subsection{Grains that strike the planet}
Inside $R_{syn}$, the surface of the planet presents a physical boundary to radial motion. Particles that strike the atmosphere are slowed and removed from orbit. The potential at the planet's surface $U(\rho,z)$ varies with latitude, and so for simplicity, we restrict our attention to planar motions where Eq.~\ref{polynomialpotential1} applies. Since, for positive grains in the equatorial plane, the potential declines as the grain moves inwards from its launch distance $r_L$, it can have at most one local maximum within $r_L$. There are thus two ways in which a grain can be prevented from striking the planet: i) the potential $U$ at the surface is greater than the launch potential, or ii) a potential peak exists between the surface and the launch position and its value is greater than or equal to the launch potential. 
\begin{figure} [placement h]
\includegraphics [height = 2.1 in] {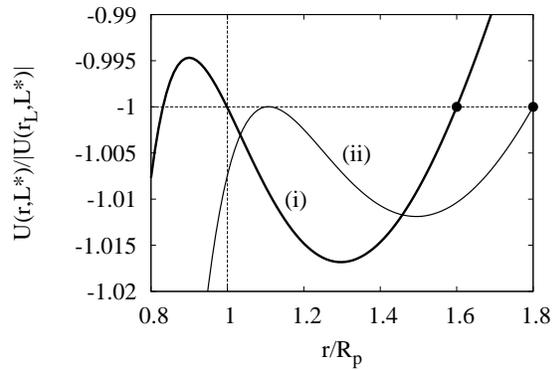}
\caption{Potential wells for two planar trajectories launched from the solid points which are inside Jupiter's synchronous orbit. Distances are in planetary radii, and the potential is scaled to the launch value. Curve (i) ($r_L = 1.6 R_p, \Lpar = 0.0991$, $a_d = 0.53 \mu$m) has a potential peak higher than the launch potential inside the planetary surface. Equating $U(r_L) = U(R_p)$ gives an analytic solution (Eq.~\ref{grazesoln}) for the stability boundary in Fig.~\ref{fig:g10JUP}b. Curve (ii) ($r_L = 1.8 R_p, \Lpar = 0.1277$, $a_d = 0.48 \mu$m) has a potential peak outside the planetary surface. In this case, the stability boundary is best obtained numerically. Both grains depicted here are poised on the stability threshold.}
\label{fig:wells2}
\end{figure}
These two scenarios are illustrated in Fig.~\ref{fig:wells2}. For case i), the stability criterion is where $U(R_p,\Lpar) = U(r_L,\Lpar)$.  Using Eq.~\ref{polynomialpotential1}, we find a quadratic expression in $\Lpar$ that implies two boundaries:
\begin{eqnarray}
\label{grazesoln}
\frac{n_L^2r_L^2}{2\Omega^2 R_p^2} \left(\frac{r_L}{R_p}-1\right)\Lpar^2+\left(1-\frac{n_Lr_L^2}{\Omega R_p^2}\right)\Lpar \nonumber &\\ +\frac12\left(\frac{r_L}{R_p}-1\right)=0.
\end{eqnarray}
The two quadratic roots of Eq.~\ref{grazesoln}, $L_1$ and $L_2$, may be obtained analytically and are plotted on Fig.~\ref{fig:g10JUP}b. The roots obey the simple expression
\begin{equation}
 L_1L_2 = \frac{r_LR_p^2}{R_{syn}^3}<1.
\label{Lroots}
\end{equation}
 Equation~\ref{Lroots} conveniently highlights several features of the lower curves in Fig.~\ref{fig:g10JUP}b: The two curves marking the grains on the threshold of collision with the planet are centered on $\Lpar<1$, as required by Eq.~\ref{Lroots}. In addition, for smaller $r_L$, the center of the instability shifts to smaller $\Lpar$, hence the left-most curve is steeper than the right-most. Finally, a planet with a larger $R_{syn}$ (eg. the Earth) will have roots that shift to very low $\Lpar$ near the planet.

The curves determined by Eq.~\ref{grazesoln} match our numerical data cleanly with two important exceptions. Firstly, because our method is only valid for grains that collide with the planet in the equatorial plane (recall our assumption $z = 0$), it misses the high latitude collisions near ($r_L = 2R_p$, $\Lpar = 2$) in Fig.~\ref{fig:g10JUP}b. All collisions exterior to the boundaries given by Eq.~\ref{grazesoln} necessarily involve substantial vertical motions, and the greyscale shading of Fig~\ref{fig:g10JUP}b shows that they do. Secondly, our criterion predicts instability for a small region near ($r_L = R_{syn}$, $\Lpar = 0.2$) that our numerical data show in fact are stable. These grains encounter a high peak, similar to curve (ii) in Fig.~\ref{fig:wells2}, that prevents them from reaching the planetary surface.  Thus $U(r_L,\Lpar) > U(R_p,\Lpar)$ is a necessary condition for radial instability in the equator plane, but it is not sufficient. 

 The additional requirement for instability is that  $U(r_{L},\Lpar) > U(r_{peak},\Lpar)$, where $r_{peak}$ is the location of an interior maximum. Just as for the escaping grains exterior to synchronous orbit, evaluation of this condition necessarily involves a cubic and a semi-analytic method. We find that no corrections to Eq.~\ref{Lroots} are needed for the high $\Lpar$ radial boundary and for all grains near the planet. Only for the right-most curve near $R_{syn}$ is there a discrepancy. Our new curve is plotted in Fig~\ref{fig:g10JUP}b and it perfectly matches the numerical instability boundary. Although the stability curve in this region can only be obtained semi-analytically, the point at which it becomes necessary occurs when the potential maximum is located at the planetary surface; $\frac{\partial U}{\partial r}\bigg{|}_{r=R_p} = 0 $  and $U(r_{L},\Lpar) = U(R_p,\Lpar)$. Evaluating these conditions, we find   
\begin{equation}
\Lpar = \frac{(\frac{r_L}{R_p}-1)^2} {\frac{R_{syn}^{3/2}r_L^{3/2}}{R_p^{3}}+2-\frac{3r_L}{R_p}}.
\label{doublesolution}
\end{equation}
For Jupiter, the critical point that satisifes both Eqs.~\ref{grazesoln} and~\ref{doublesolution} is at $\Lpar  = 0.112, r_L=1.694 R_p$ (solid point in Fig~\ref{fig:g10JUP}b). The stability curve meets $r_L=R_{syn}$ at $\Lpar = 2-\sqrt{3}$, a result suggested by our local stability analysis of section 3. Note that our energy arguments yield analytic expressions both inside and outside $R_{syn}$. Arguments involving the location of potential maxima, conversely, require semi-analytic methods.

\begin{figure}[placement h]
\vspace{0.2 in}
\includegraphics [height = 2.1 in]{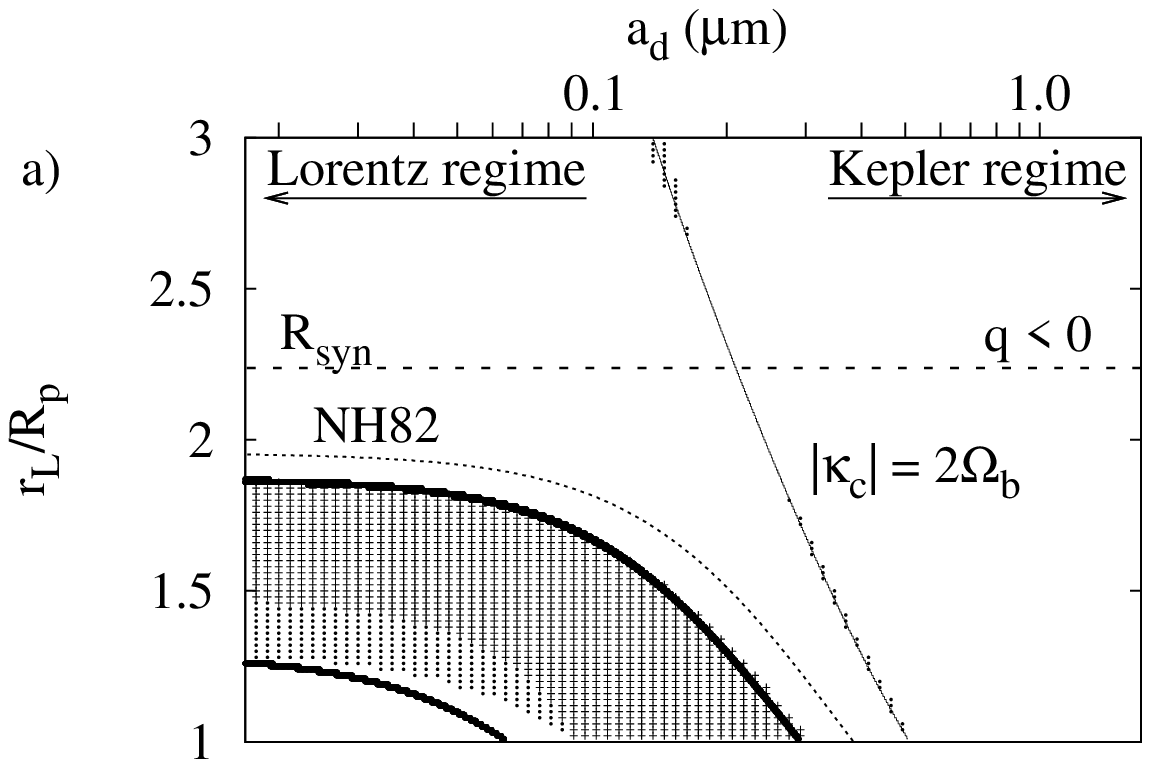}
\newline
\includegraphics [height = 2.1 in]{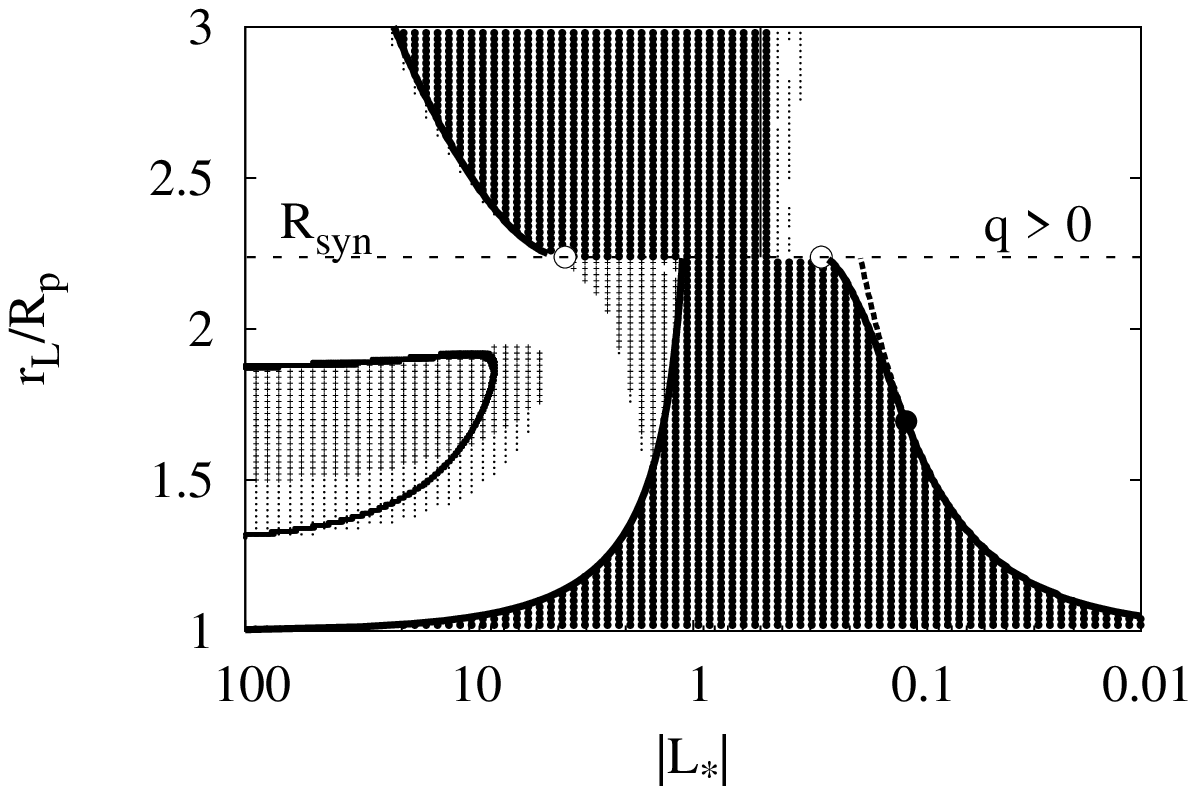}
\caption{Our new analytic results (heavy solid lines) are plotted over the numerical data from Fig.~\ref{fig:g10num}. \textbf{a)} Northrop's solution (dotted line) is superseded by our two semianalytic boundaries where  $\langle\frac{\partial^2 U}{\partial z^2}\rangle = \Omega_b^2 = 0$ from  Eq.~\ref{dphidz}. The new boundaries are a significantly better fit to the data and indicate an inner stability zone. The $|\kappa_c| = 2\Omega_b$ curve indicates the 2:1 resonance between the epicyclic and the vertical bounce frequencies; it matches the data points well. \textbf{b)} We extend our vertical stability boundary to positive grains. The radial stability boundaries for grains that escape or crash into the planet are discussed in the text (section 4). Between the open circles at $r_L = R_{syn}$ and $\Lpar = 2 \pm \sqrt{3}$, orbits are locally, radially unstable. The solid circle is the critical point defined by Eqs.~\ref{grazesoln} and~\ref{doublesolution}.}
\label{fig:g10JUP} 
\end{figure}
\section{Local Vertical Stability Analysis}
The stability of grains against vertical perturbations was first explored by \citet{nh82}. In their model, a grain is launched on a circular orbit at the equilibrium orbital frequency $\omega_c$ in the potential of Eq.~\ref{potentialcylinder} so that there is no gyromotion around magnetic field lines. If the grain orbit at the equilibrium point is stable to vertical perturbations, the square of the bounce frequency $\Omega_b$, given by
\begin{eqnarray}
\Omega_b^2  =  \frac{\partial^2U}{\partial z^2}\bigg{|}_{\rho= \rho_c, z=0} =  \frac{G M_p}{\rho_c^3}\left(\frac{3\Lpar\dot{\phi_c}}{\Omega}+ 1\right) \nonumber &\\  = 3\omega_c^2-2n_c^2
\label{nh82e7}
\end{eqnarray} 
\citep{nh82} is positive. Here $\dot{\phi_c} = \omega_c -\Omega$ is the dust grain's azimuthal frequency in the frame rotating with the magnetic field. When multiplied by $z$, Eq.~\ref{nh82e7} gives the centrifugal (first term), and gravitational (last term) accelerations along a magnetic field line.

For $\Omega_b^2 < 0$, the vertical motion is unstable. Note that the gravitational acceleration is negative and thus destabilizing. This follows from the fact that the dipolar magnetic field curves toward the planet and so a grain leaving the equatorial along a field line plane moves downhill in the gravitational potential. The \citet{nh82} solution for the boundary where $\Omega_b = 0$ is plotted in Figs.~\ref{fig:g10num}a and~\ref{fig:g10JUP}a. At distances closer to the planet than a critical distance $\rho_{crit}$, gravity forces grains to leave the equatorial plane.
\subsection{Vertical Instability in the Lorentz Limit}
In the limit of high charge-to-mass ratio, Eq.~\ref{nh82e7} can be solved exactly:
\begin{equation}
\label{nh82rhocrit}
\frac{\rho_{crit}}{R_{syn}} = (2/3)^{\frac13} \approx 0.87.
\end{equation}
The effect of our initial condition, launching grains at the Kepler speed, however, necessarily causes epicyclic gyromotion as the grain orbits the planet. This leads to a stabilizing magnetic mirror force, in which the grain resists moving out of the equatorial plane to regions of higher magnetic field strength as discussed in section 1.1. Following the procedure of \citet{lew61} and \citet{tva80}, the magnetic mirror force for equatorial pitch angles near 90$^{\circ}$ adds a component of strength $9r_g^2\Omega_{gc}^2/2\rho_c^2$ to Eq.~\ref{nh82e7}. In the Lorentz limit, Eq.~\ref{gyroradius} simplifies to $r_g \Omega_{gc} = \rho_c(\Omega-n_c)$, and the bounce frequency can be found from:
\begin{equation}
 \Omega_b^2 =  3\Omega^2 - 2n_c^2 + \frac{9}{2}(n_c - \Omega)^2.
\label{OmbhighL}
\end{equation}
 As above, the first two terms are due to the centrifugal and gravitational forces on a grain tied to a nearly vertical magnetic field line. The third term of Eq.~\ref{OmbhighL} is the magnetic mirror term, generalized to account for a rotating magnetic field. The three vertical accelerations add linearly, and are valid in the limit that $\Lpar \rightarrow \infty$ and $r_g \rightarrow 0$.
\begin{figure}[placement h]
\includegraphics [height = 2.1 in]{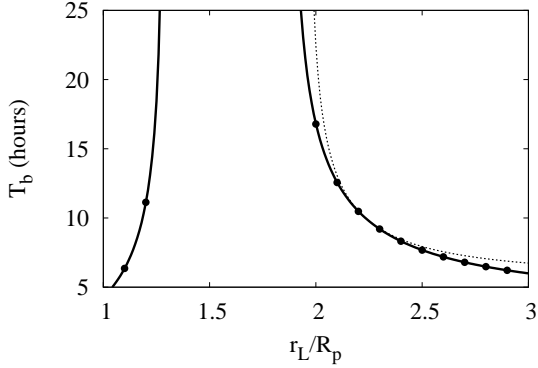}
\caption{The bounce period for $\Lpar=-10^{4}$ grains at Jupiter over a range of launch distances.  Northrop's solution (Eq.~\ref{nh82e7}, dotted line) and our solution (Eq.~\ref{dphidz}, solid lines) with $T_b = 2\pi/\Omega_b$, are plotted alongside numerical data (points). Note that in our solution $T_b$ is smaller than in Northrop's solution everywhere except at $R_{syn} = 2.24 R_p$. In the limit where $r_c << R_{syn}$, Eq.~\ref{OmbhighL} shows that grains satisfy $\Omega_b \rightarrow \sqrt{\frac{5}{2}}n_c$, while for $r_c >> R_{syn}$, $\Omega_b \rightarrow \sqrt{\frac{15}{2}}\Omega$ and $T_b \rightarrow 3.62$ hours.}
\label{fig:Tbounce} 
\end{figure}

Fig.~\ref{fig:Tbounce} compares the \citet{nh82} bounce period (Eq.~\ref{nh82e7}) with our Eq.~\ref{OmbhighL} which accounts for epicyclic motion for small dust grains at Jupiter. The Northrop formalism erroneously predicts bounce periods that are too long both inside and outside synchronous orbit and, more seriously, misses the second solution near the planet.

The third term in Eq.~\ref{OmbhighL} is positive everywhere inside the Northrop boundary and thus leads to enhanced vertical stability. The stability boundaries in the high $|\Lpar|$ limit are determined by Eq.~\ref{OmbhighL}; setting $\Omega_b = 0$, we find:
\begin{equation}
\label{rhocrit2}
\frac{r_L}{R_{syn}} = \left(\frac{5}{9 \pm \sqrt{6}}\right)^{2/3} \approx 0.58, 0.84 .
\end{equation}
These limits are valid for both positive and negative grains with $|\Lpar| \rightarrow \infty$.  Between these limits, $\Omega_b^2 < 0$ and grain orbits are locally unstable; the enhanced stability from the mirroring force moves the vertical stability boundary inwards from Northrop's 0.87$R_{syn}$ to 0.84$R_{syn}$. A more important change, regained stability inside 0.58 $R_{syn}$, is due to the higher launch speeds relative to the field lines, larger gyroradii, and a stronger magnetic mirror force. For Jupiter these distances are at $1.29 R_p$ and $1.87 R_p$ respectively (see Fig.~\ref{fig:g10JUP}). Hints of this inner stability zone were seen numerically by \citet{nh83a} and \citet{nc87}; here we have derived analytical solutions for vertical stability in the Lorentz limit.

\subsection{Vertical instability for all charge-to-mass ratios}
To extend our model for bounce motion over all charge-to-mass ratios we must, in principle, account for the variation in the strengths of the vertical gravitational, centrifugal and electromagnetic accelerations over one gyrocycle. Extending the electromagnetic mirror acceleration requires breaking the assumption of perfectly circular gyrocycles, and is beyond the scope of this work. The remaining two accelerations, however, can be extended to second order in $r_g/\rho_c$ while retaining circular gyrations. We begin by writing the vertical acceleration as a function of the epicyclic phase $\theta$:
\begin{eqnarray}
\frac{\partial^2U}{\partial z^2}z =  \frac{G M_p z(\theta)}{\rho^3(\theta)}\left(\frac{3\Lpar\dot{\phi}(\theta)}{\Omega}+ 1\right).
\label{nh82e7theta}
\end{eqnarray}
To first order in $r_g$, the epicycles are circles in the guiding center frame. Setting $\theta = 0$ at the closest point to the planet, we find
\begin{equation}
\rho(\theta) = \rho_c - |r_g| \cos{\theta},
\label{rho(theta)}
\end{equation}
and
\begin{equation}
\dot{\phi}(\theta) = \dot{\phi_c} - \kappa_c \frac{|r_g|}{\rho_c}\cos{\theta}.
\label{phidot(theta)}
\end{equation}
Due to the geometry of a dipole near its equator, an epicycle is tilted by an angle $\approx 3 \lambda$ (where $\lambda$ is the latitude). Hence the vertical offset is given by:
\begin{equation}
z(\theta) = z_c - \frac{3|r_g|}{\rho_c}\cos{\theta},
\label{z(theta)}
\end{equation}
To calculate the bounce frequency, we average the restoring acceleration over an epicycle, a procedure that is valid as long as $\kappa_c >> \Omega_b$:
\begin{equation}
 \Omega_b^2 =  \frac{\biggr\langle z \frac{\partial^2 U}{\partial z^2} \biggr\rangle}{\langle z \rangle} = \frac{1}{2\pi z_c}\int^{2\pi}_{0}\frac{\partial^2 U}{\partial z^2}z(\theta) d\theta.
\label{gyroaverage}
\end{equation}
Using Eqs.~\ref{rho(theta)} and~\ref{phidot(theta)} to eliminate $\rho(\theta)$ and $\dot{\phi}(\theta)$ in Eq.~\ref{nh82e7theta}, we expand to $O(r_g^2)$, integrate Eq.~\ref{gyroaverage}, and add in the magnetic mirroring term from Eq.~\ref{OmbhighL} to obtain:
\begin{eqnarray}
\label{dphidz}
 \Omega_b^2 =  3 \omega_c^2-2n_c^2+\frac{9}{2}(n_c-\Omega)^2 \nonumber &\\ -\frac{r_g^2}{\rho_c^2}\left(\frac{9}{2}\Omega_{gc}\dot{\phi}_c +\frac{3}{2}n_c^2 \right).
\end{eqnarray}
The frequencies in Eq.~\ref{dphidz}, $\omega_c$ (Eq.~\ref{dUdr}), $n_c$ (Eq.~\ref{nkepL}), $\Omega_{gc}$ (Eq.~\ref{gyrofrequencyL}), and $\dot{\phi}_c = \omega_c - \Omega$, are all evaluated at the guiding center of motion $\rho_c = r_L + r_g$, which is determined by Eq.~\ref{gyroradius}. Our calculation adds two additional destabilizing terms that are strongest for intermediate values of $\Lpar$, where gyroradii are largest (Fig.~\ref{fig:radrange}).

How does our solution compare to numerical data? In Fig.~\ref{fig:g10JUP}, we plot our theoretical curves against the numerical data for both negative and positive grains launched at the Kepler rate in an aligned dipole field for Jupiter. We find the curves tracing the unstable zone semi-analytically by setting $\Omega_b = 0$ in Eq.~\ref{dphidz}. Within the regions bordered by the curves, trajectories are locally unstable but may remain globally bound due to high-latitude restoring forces.   

Our model closely matches the outer stability boundary for negative grains but is less successful for the inner boundary, especially for moderate $\Lpar$. This is precisely where our derivation is weakest; recall that we have not accounted for higher-order corrections for the magnetic mirror force which are strongest closest to the planet and for $|\Lpar| \sim 1$. Near $|\Lpar| = 1$ epicycles become large and distorted for negative grains and even more so for positive grains (\citealt{mhh82,mhh03}). Figure~\ref{fig:radrange}b shows that the epicyclic model matches the radial range of positively-charged grains well for values $\Lpar>10$. This is exactly where the numerical data depart from the theory in Fig.~\ref{fig:g10JUP}b. Apparently, large gyroradii and interference from the proximate radial instability strip lead to unmodeled effects and excess vertical instability.

The curvature of the outer boundary in Fig.~\ref{fig:g10JUP}a is similar to that for the Northrop instability, albeit displaced to locations closer to the planet. Notice that, with decreasing $|\Lpar|$, the instability region curves towards the planet for negative grains, and away from it for positive grains (Fig.~\ref{fig:g10JUP}). This is primarily due to the $3\omega_c^2-2n_c^2$ term that determines the Northrop boundary. For negative grains inside synchronous orbit, $n_c > \omega_c$, and $\omega_c$ increases with decreasing $|\Lpar|$ due to a weakening outwardly-directed electromagnetic force. It thus takes a greater value of $n_c$ to make $3\omega_c -2n_c$ change sign, destabilize the vertical motion, and move the boundary curves to lower launch distances in Fig.~\ref{fig:g10JUP}a. For the positive grains in the Lorentz limit, by contrast, $\omega_c$ decreases as $\Lpar$ decreases, and a smaller $n_c$ will destabilize the grain. Thus with decreasing $\Lpar$ this boundary in Fig.~\ref{fig:g10JUP}b curves up to higher launch radii. 

Finally, notice the band of locally unstable but globally stable points that stretches from $|\Lpar| \approx 0.1$ at the surface of the planet to $|\Lpar| \approx 1$ at large distances in Figs.~\ref{fig:g10num}a and ~\ref{fig:g10JUP}a. These grains are affected by a $|\kappa_c| = 2\Omega_b$ resonance that couples their radial and vertical motions. Energy is transferred from the radial oscillation to a vertical oscillation and back again. Near the synchronous orbit, gyroradii are initially small and therefore there is not as much radial motion to transform into vertical motion; these grains do not reach our $\lambda_m = 5^{\circ}$ threshold and appear as white space in Fig.~\ref{fig:g10JUP}a. 
 
The existence of stable trajectories within the Northrop boundary is an important result, particularly for small slowly-rotating planets with distant synchronous orbits like Earth. Small dust grains generated by the collisional grinding of parent bodies on Keplerian orbits can remain in orbits near the planetary surface. High energy plasma, like that found in Earth's van Allen radiation belts, is more stable than we have calculated here by virtue of exceedingly rapid gyrations and a greatly enhanced mirroring force.
 
Our analysis to this point is completely general and, although we have focused on Jupiter, can be easily applied to other planets. Saturn and Earth are logical choices, as their magnetic fields are also dominated by the $g_{10}$ aligned dipolar component. The appearance of the stability map for any planet depends on only the parameters $R_{syn}$ and $R_p$, and not on the substantially different magnetic field strengths which, due to our use of $\Lpar$, only affect the conversion to grain radius $a_d$.  The synchronous orbital distance is somewhat closer to the planetary surface at Saturn ($R_{syn} = 1.86 R_p$) than at Jupiter ($R_{syn} = 2.24 R_p$), while at Earth ($R_{syn} = 6.61R_p$) it is much further away. This leads to interesting differences between the planets, as we shall see below.
\section{Saturn and Earth}
\begin{figure} [placement h]
\vspace {0.1 in}
\includegraphics [height = 2.1 in] {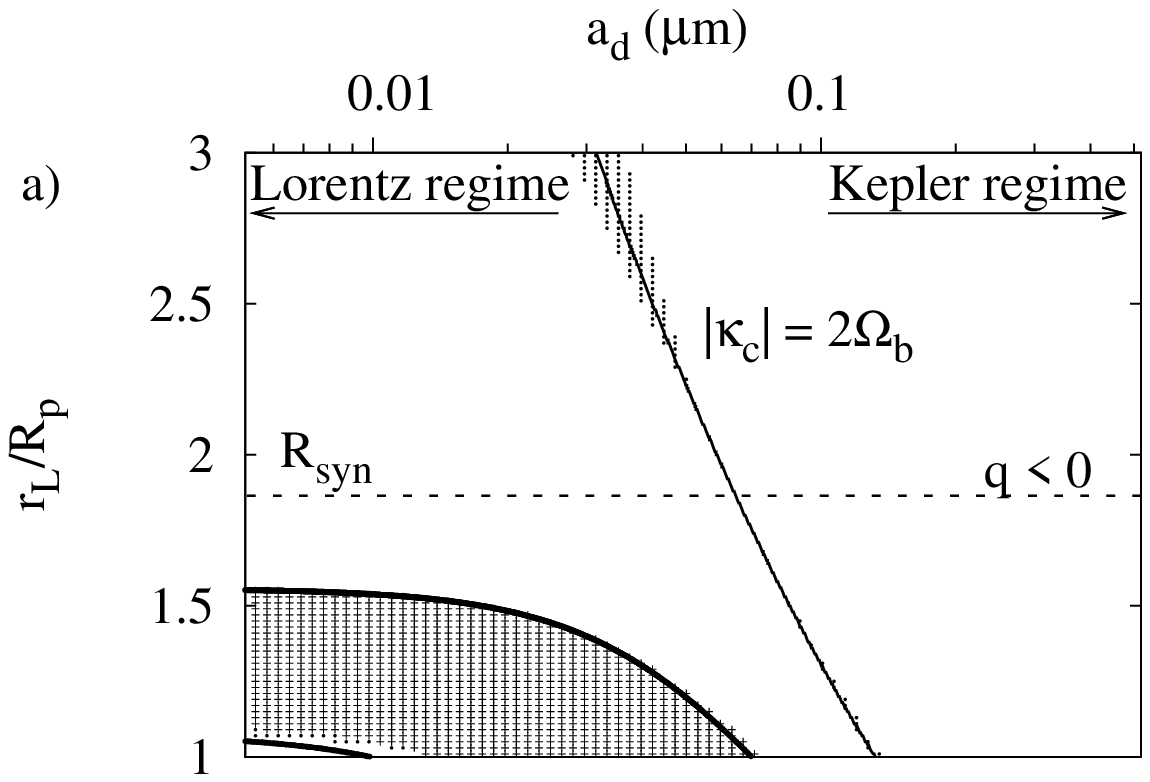}
\newline
\includegraphics [height = 2.1 in] {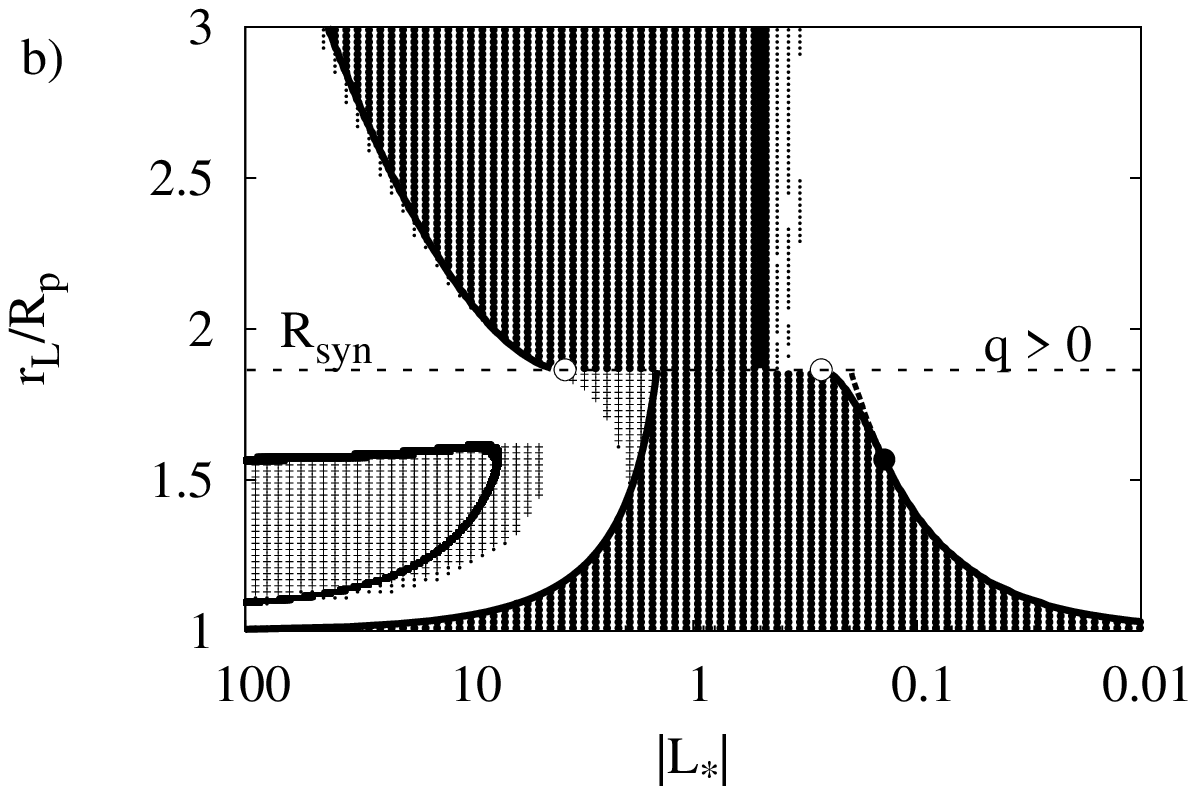}
\caption{Stability of charged grains at Saturn modeled with a centered and aligned dipole field. All initial conditions and theoretical curves are as in Fig.~\ref{fig:g10JUP}. Also as in Fig.~\ref{fig:g10JUP}, the darkest shade of grey signifies low latitude collision or escape, the middle shade indicates high latitude collisions, and the lightest grey signifies large vertical excursions. For negative charges (panel \textbf{a}) only a tiny stable region exists near ($r_L = R_p$, $\Lpar = -50$) due to Saturn's smaller $R_{syn}$. Furthermore, due to the proximity to Saturn, nearly all grains that are locally vertically unstable do in fact hit the planet, unlike their counterparts at Jupiter. \textbf{b)} positive charges. As with the negative grains, nearly all the vertically unstable grains hit Saturn. Saturn's radial instability region (darkest grey) looks much like Jupiter's.}
\label{fig:SATg10}
\end{figure}

A centered and aligned dipole is an excellent approximation for Saturn's magnetic field. We take $g_{10} = 0.2154$ Gauss from \citet{cdc84} and plot both our numerical data and analytical stability boundaries in Fig.~\ref{fig:SATg10}. The Cassini measurement of $g_{10}$ does not vary significantly from the older value that we use \citep{bdr09}.
A lower synchronous orbit at Saturn pushes the local vertical instability inward, as expected from Eq.~\ref{rhocrit2}. Comparing Fig.~\ref{fig:SATg10} to Fig.~\ref{fig:g10JUP}, we see that the proximity of the surface at Saturn causes all the locally vertically unstable grains to physically collide with the planet. This is true for both negative and positive grains.

Outside synchronous orbit in Fig.~\ref{fig:SATg10}b, the solutions derived for positive escaping grains in section 4 apply at Saturn to very high accuracy, for both the low $\Lpar$ and high $\Lpar$ boundaries. As in Fig.~\ref{fig:g10JUP}b, grains with $\Lpar \lesssim \frac12$, do not have enough energy to escape despite achieving large radial excursions (light grey region outside $R_{syn}$ in Fig.~\ref{fig:SATg10}b). For these grains, vertical motions are excited over several orbits, as in Fig.~\ref{fig:JUPg10randomwalk}.

 Within synchronous orbit, the condition $U(R_p) = U(r_L)$ (solved in Eq.~\ref{grazesoln}) bounds most of the unstable grains. As at Jupiter, a small set of large grains near $R_{syn}$ require the semi-analytical analysis of the potential between the launch position and the surface to determine global stability. This analysis yields the curve connecting the filled black circle at ($r_L = 1.568 R_p$, $\Lpar =0.14 $) and the open circle ($r_L=R_{syn}$, $\Lpar = 2-\sqrt{3}$) in Fig.~\ref{fig:SATg10}b.

 Compared to Jupiter and Saturn, Earth's magnetic field is ``inverted'' at the current epoch, with magnetic north near the geographic south pole ($g_{10} = -0.3339$ Gauss taken from Roberts and Soward (1972). Thus at Earth, $\Lpar > 0$ for negative grains. This causes positive grains to be radially stable, gyrating between the launch position and synchronous orbit, and negatively-charged grains to be radially unstable. The Earth is also far smaller on the scale of its own synchronous orbit than the gas giants, and so serves as an excellent test of the accuracy of our analytical solutions far from $R_{syn}$.
\begin{figure} [placement h]
\vspace {0.1 in}
\includegraphics [height = 2.1 in] {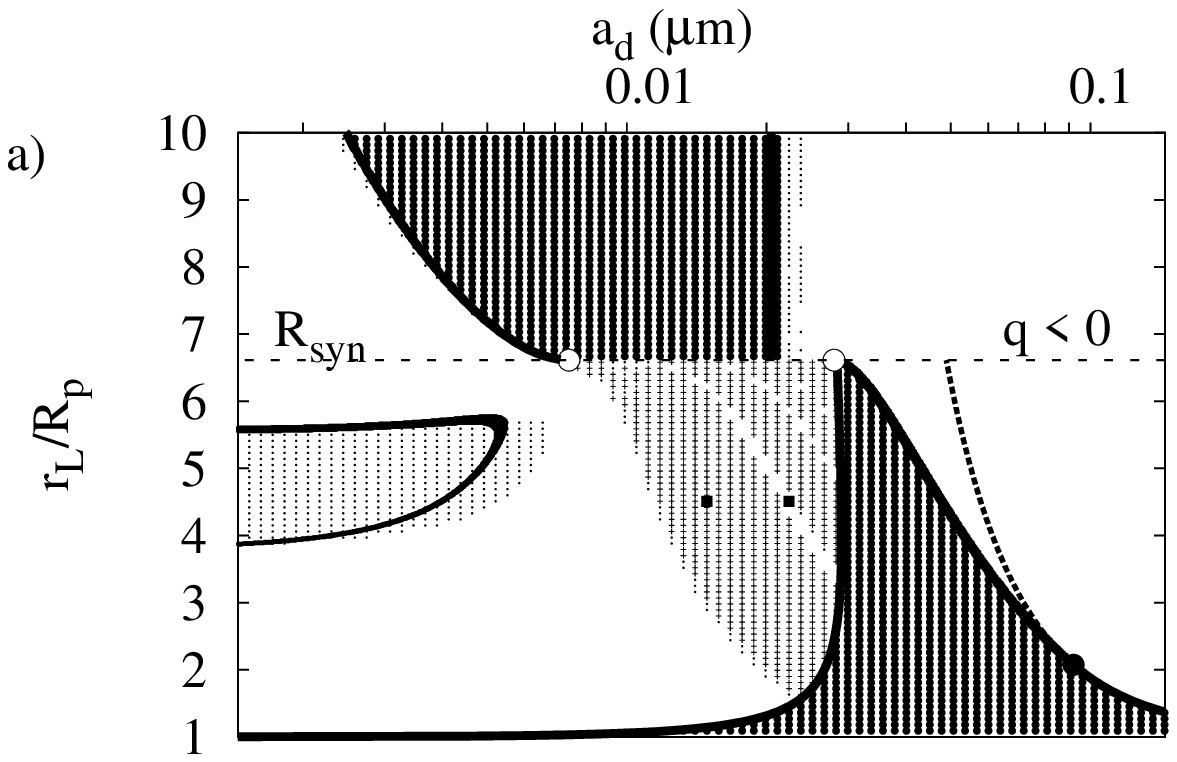}
\newline
\includegraphics [height = 2.1 in] {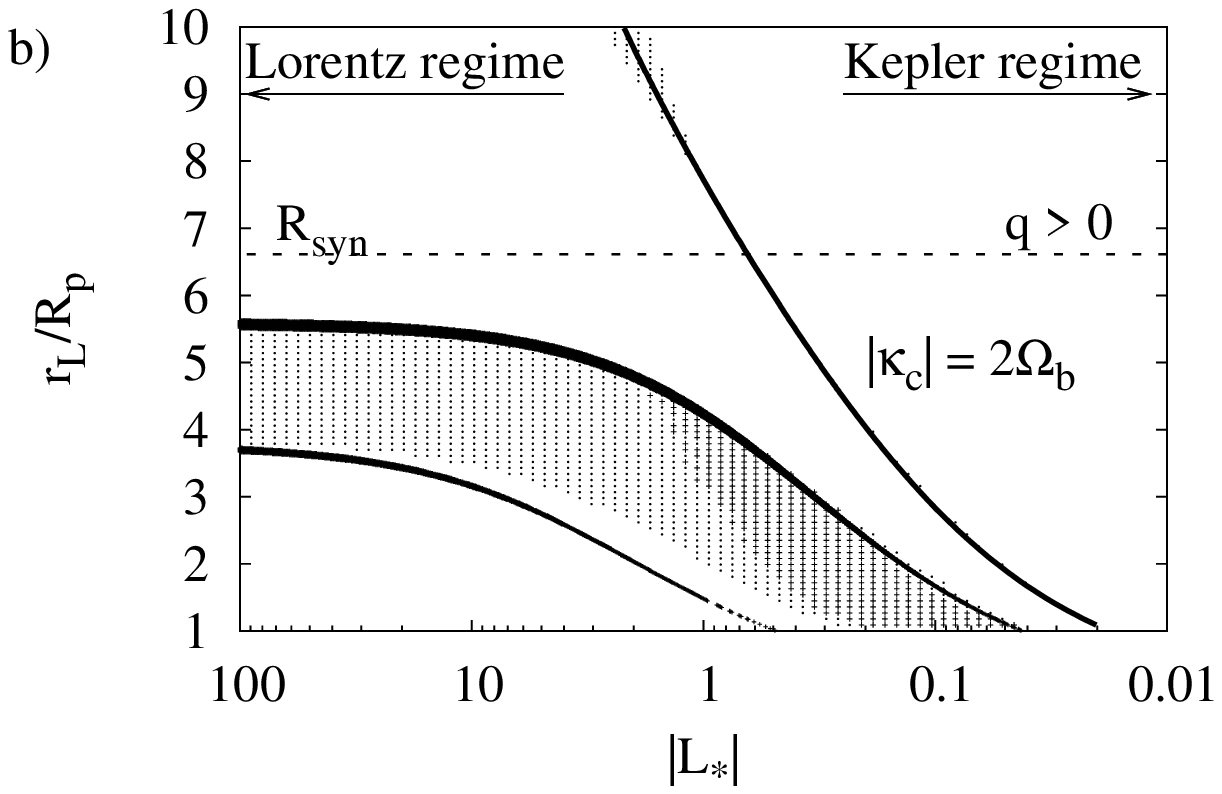}
\caption{Stability of charged grains at Earth, modelled with a centered and anti-aligned dipole field. Theoretical curves and initial conditions are the same as in Figs.~\ref{fig:g10JUP} and ~\ref{fig:SATg10}. Since $R_{syn}$ is much larger than for Jupiter and Saturn, we extend the radial range of the integrations to $r_L = 10R_p$ and the distant threshold signifying escape to $r_{esc} = 100 R_p$. The open circles at ($r_L = R_{syn}$, $\Lpar = 2+\sqrt{3}$) and ($r_L = R_{syn}$, $\Lpar = 2-\sqrt{3}$) are as in Fig.~\ref{fig:g10JUP}, and the solid circle, marking the transition from the analytical to semi-analytical boundary for the larger grains is at $\Lpar= 0.0248, r_L = 2.074R_p$. The two solid squares in \textbf{a)} are individual grain trajectories illustrated in Figs.~\ref{fig:EARtraj4} and~\ref{fig:EARstabletrack}. }
\label{fig:EARg10}
\end{figure}
For the Earth, Fig.~\ref{fig:EARg10}a shows the radial global instabilities. Outside $R_{syn}$, the boundaries are in excellent agreement with our analytical results for large and small grains. Inside $R_{syn}$, grains are globally radially unstable and all the grains that collide with the planet at low latitudes are launched between our two solutions given by Eq.~\ref{grazesoln}. The set of grains for which Eq.~\ref{grazesoln} is an insufficient criterion for collision with the planet, however, is much larger at the Earth than at Jupiter or Saturn. For Earth, just like for the gas giants, the global solution for radial stability inside synchronous orbit perfectly matches the numerical data and meets $R_{syn}$ at the local stability solution ($r_L=R_{syn},\Lpar=2-\sqrt{3}$). Furthermore, the solutions of Eq.~\ref{grazesoln} have shifted to much lower $\Lpar$ (see Eq.~\ref{Lroots}), reducing the total range in $\Lpar$ for grains which collide with the planet at low latitude.

The local vertical stability boundary matches the numerical data well, although in the Lorentz limit, all grains are globally stable since the high latitude restoring forces become much stronger close to the planet \citep{hdh00}. Only at $|\Lpar| \leq 1$ do the positive grains collide with the planet. As in Figs.~\ref{fig:g10JUP} and~\ref{fig:SATg10} the vertical stability curves match very well for large $\Lpar$ and deviate from the data for $\Lpar \approx 1$. The $|\kappa_c| = 2\Omega_b$ resonance also matches the data well.

\begin{figure} [placement h]
\includegraphics [height = 2.1 in] {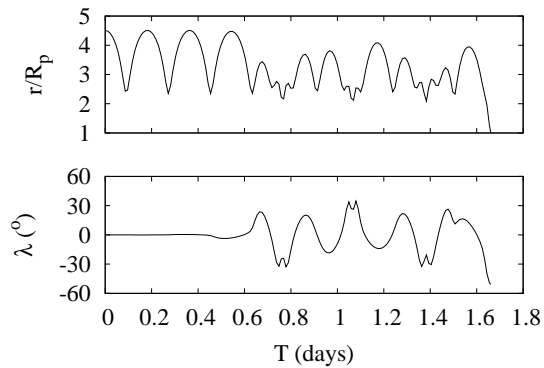}
\caption{A grain with large radial excursions that gradually excite substantial vertical oscillations at the Earth, ($r_L = 4.51R_p$, $\Lpar = 0.948$, $a_d = 0.0149\mu$m).}
\label{fig:EARtraj4}
\end{figure}

Earth has a much larger class of grains that experience large radial excursions, which excite vertical motions. Most of these grains, from the medium-grey areas on the stability map of Fig.~\ref{fig:EARg10}a that link the disjoint dark grey regions of global radial instability, collide with the planet at high latitudes. An example of a trajectory in this class is shown in Fig.~\ref{fig:EARtraj4}.

\begin{figure} [placement h]
\includegraphics [height = 2.1 in] {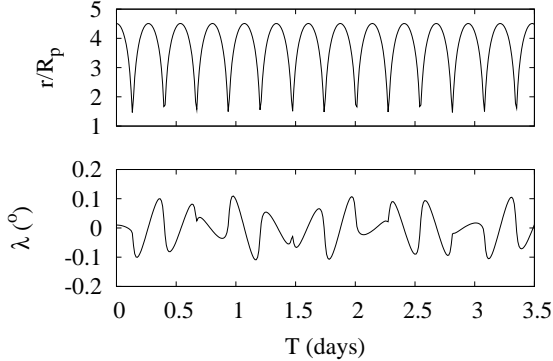}
\caption{A grain with large radial motions like that depicted in Fig.~\ref{fig:EARtraj4} that nevertheless always remains near the equatorial plane ($r_L = 4.51R_p$, $\Lpar = 0.419$, $a_d = 0.0224\mu$m).}
\label{fig:EARstabletrack}
\end{figure}

At Saturn all of the grains in this region collided with the planet, but at the Earth we see three white tracks of orbits that never leave the equatorial plane, and hence are energetically prevented from striking the planet. We plot an example in Fig.~\ref{fig:EARstabletrack}. A few of these trajectories are also apparent for Jupiter (Figs.~\ref{fig:g10num}b and~\ref{fig:g10JUP}b.) We suspect, based on the similarity of the white stable tracks in Fig.~\ref{fig:EARg10}a and the $|\kappa_c| = 2\Omega_b$ line in Fig.~\ref{fig:EARg10}b, that these are resonant phenomena.
\section{Conclusion}
For Kepler-launched grains in centered and aligned dipole planetary fields, we have employed both local and global stability analyses to provide solutions for stability boundaries that match numerical simulations for Jupiter, Saturn and the Earth.
\begin{figure} [placement h]
\includegraphics [height = 2.1 in] {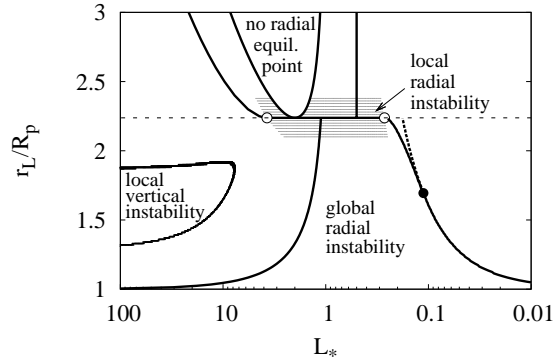}
\caption{Local and global stability boundaries for positive grains at Jupiter. The dashed line is $R_{syn}$, and the shaded region highlights local radial instability near $R_{syn}$. The white circles bounding this region on $R_{syn}$ ($\Lpar = 2 \pm \sqrt{3}$ for all planets) also show where our semi-analytical curves for grains that escape or collide with the planet meet $R_{syn}$. The solid circle is where the potential at the planet's surface is a local maximum and equals to the launch potential.}
\label{fig:JUPlocalanalysis}
\end{figure}
 Figure~\ref{fig:JUPlocalanalysis} provides a summary of the various analytical results discussed in this work for positive grains at Jupiter.

We find that local radial stability is very useful in the immediate vicinity of synchronous orbit, since $r_g \rightarrow 0$ there (Eq.~\ref{gyroradius}). More importantly, our restriction of the global radial analysis to equatorial orbits is justified by the excellent agreement between analytics and numerics. Radial instability has important implications for depleting particles near the surface of a planet but beyond the reach of atmospheric drag forces. At Earth, for example, the radial instability eliminates negatively-charged particles with $r_g \lesssim 0.2\mu$m from Low Earth orbit, and $\lesssim 0.1 \mu$m from within 2000 km. For Jupiter, this instability sweeps positive grains with $r_g < 1\mu$m from the region within 10,000 km from Jupiter's cloud-tops.

 Our local vertical analysis of grains launched on Kepler circles in the equatorial plane adds the effect of the magnetic mirror force and is a major improvement to the equilibrium model of \citet{nh82}. We do not undertake a fully global analysis which would seek to distinguish grains that strike the planet from those that simply sustain large amplitude oscillations in latitude. 

Although the magnetospheres of Jupiter, Saturn and the Earth are all nearly dipolar, each planet has additional components that make the field more complicated. Saturn has the simplest field and is well represented by a dipole offset northward by a few thousand km. Jupiter and the Earth have non-zero dipole tilts that cause the magnetic field seen by an orbiting grain to fluctuate. Nevertheless, since tilts and offsets are generally small, we expect that the radial forces will be only slightly affected, and the radial instability region will remain nearly the same. Vertical motions, by contrast, should be strongly affected since a circular orbit in the equatorial plane is no longer an equilibrium point. The global radial analysis, which included the effects of radial oscillations, led to a much larger instability region than the simple local analysis (top of Fig.~\ref{fig:JUPlocalanalysis}); in exactly the same way, we expect the region of vertical instability to expand substantially when dipole tilts or offsets are included. We will take up the study of more complicated magnetic field configurations and launch conditions in a forthcoming paper.

\end{document}